\documentclass[%
 reprint,
 superscriptaddress,
 amsmath,amssymb,
prl,
10pt
]{revtex4-2}

\usepackage[colorlinks=true,linkcolor=blue,citecolor=blue]{hyperref}
\usepackage{bm}

\usepackage[english]{babel}
\usepackage{microtype}
\usepackage{indentfirst}
\usepackage{graphicx}
\usepackage{placeins}
\usepackage{siunitx, booktabs}
\usepackage{tabularx}
\usepackage[caption=false]{subfig}
\usepackage{amsmath}
\usepackage{array}
\usepackage{dcolumn}
\usepackage{bm}
\usepackage{setspace}
\usepackage{amssymb}
\usepackage{braket}
\usepackage{xcolor}
\usepackage{color,soul}
\usepackage{placeins}
\usepackage{eqnarray,amsmath}

\newcommand{\FIG}[1]{\textcolor{red}{Fig. }}
\newcommand{\TAB}[1]{\textcolor{violet}{Table }}

\begin{document}

\title{Efficient Full-frequency GW Calculations using a Lanczos Method}

\author{Weiwei Gao}
\address{Key Laboratory of Materials Modification by Laser, Ion and Electron Beams, Ministry of Education, Dalian University of Technology, Dalian 116024, China}

\author{Zhao Tang}
\address{Center for Computational Materials, Oden Institute for Computational Engineering and Sciences, The University of Texas at Austin, Austin, TX 78712}

\author{Jijun Zhao}
\email{zhaojj@dlut.edu.cn}
\address{Key Laboratory of Atomic and Subatomic Structure and Quantum Control (Ministry of Education), Guangdong Basic Research Center of Excellence for Structure and Fundamental Interactions of Matter, School of Physics, South China Normal University, Guangzhou 510006, China}

\author{James R. Chelikowsky}
\email{jrc@utexas.edu}
\address{Center for Computational Materials, Oden Institute for Computational Engineering and Sciences, The University of Texas at Austin, Austin, TX 78712}
\address{Department of Physics, The University of Texas at Austin, Austin, TX 78712}
\address{McKetta Department of Chemical Engineering, The University of Texas at Austin, Austin, TX 78712 }

\begin{abstract}
The GW approximation is widely used for reliable and accurate modeling of single-particle excitations. It also serves as a starting point for many theoretical methods, such as its use in the Bethe-Salpeter equation (BSE) and dynamical mean-field theory.
However, full-frequency GW calculations for large systems with hundreds of atoms remain computationally challenging, even after years of efforts to reduce the prefactor and improve scaling. 
We propose a method that reformulates the correlation part of the GW self-energy as a resolvent of a Hermitian matrix, which can be efficiently and accurately computed using the standard Lanczos method. 
This method enables full-frequency GW calculations of material systems with a few hundred atoms on a single computing workstation. 
We further demonstrate the efficiency of the method by calculating the defect-state energies of silicon quantum dots with diameters up to 4 nm and nearly 2,000 silicon atoms using only 20 computational nodes. 
\end{abstract}

\maketitle


As a first-principles approach based on many-body perturbation theory, the GW approximation has been successfully applied to accurately compute quasiparticle excitation in weakly and moderately correlated materials~\cite{hedin1965,hybertsen1986,godby1988}. 
The approximation also plays an essential role in the first-principles calculations of excitonic effects using the GW+BSE approach~\cite{rohlfing1998,Albrecht1998} and is used in conjunction with other methods~\cite{yanghao2021,Biermann2003,Ping2002,Tianyu2021,zhenglu2019,zhenglu2021}. 
The computational scaling of different implementations of GW approximation ranges from $O(N)$ to $O(N^6)$ and typically has a much larger pre-factor compared to density functional theory (DFT) calculations with semi-local exchange-correlation functionals~\cite{golze2019, Gao_2022}. 

During the last decade, different formulations and algorithms have been proposed and implemented for accelerating GW calculations to meet the challenge of modeling large and complex materials~\cite{golze2019,Gao_2022}. 
A few seminal papers have demonstrated GW calculations of quasiparticle energies of large systems with the number of atoms ranging from 1,000 to around 2,700~\cite{wilhelm2018,vlcek2018,Mauro2020,duchemin2021_cubicscale,victor2022}. 
These large-scale GW calculations rely on well-crafted numerical optimization and large computation resources, which are of limited accessibility. 
Even with notable advancements, GW calculations for systems with a few hundred atoms, which are typically required for computationally studying a point defect in solids or small quantum dots, cannot be performed routinely. 
Owing to the significant expense of data curation, GW calculations are rarely used in wide-reaching data-driven research, such as constructing large databases of material properties and training supervised machine-learning models. 

In GW calculations, one of the most computationally expensive steps is calculating the frequency-dependent screened Coulomb potential $W$. 
In many implementations, the irreducible polarizability function and then the inverse dielectric function are calculated to compute $W$~\cite{DESLIPPE20121269,Sangalli_2019}. 
Such a procedure deals with the frequency dependence of $W$ using approximations like plasmon-pole models or numerical tools such as contour deformation or analytical continuations~\cite{golze2019,Hedin_1999}. 
Alternatively, another approach computes the reducible polarizability and $W$ by solving the Casida equation derived in linear-response time-dependent density functional theory~\cite{casida1995,BRUNEVAL2016149,Tiago2006,vansetten2013,daniel2021}. 
Once all the eigenvalues and eigenvectors of the Caisda equation are solved, the frequency-dependent $W$ and GW quasiparticle self-energies can be written and computed in a closed form~\cite{BRUNEVAL2016149,Tiago2006,vansetten2013}. 
While this approach is formally simple and works efficiently for small systems with less than 40 atoms, it becomes numerically intractable for large systems due to the high cost of solving the Casida equation. 


Here, we propose a method that avoids solving the Casida equation while still allowing us to perform full-frequency GW calculations analytically and efficiently. 
To better illustrate the concept, we discuss our method applied to finite systems, for which real-valued wave functions and simplified notations can be used.  
Initially, we perform DFT calculations to obtain Kohn-Sham orbitals $|\phi_m \rangle $ and their corresponding energies $\epsilon_m$, which are used as an initial approximation for quasiparticle wave functions and energies, respectively. 
Next, the Casida equation can be constructed~\cite{casida1995,Tiago2006,BRUNEVAL2016149,daniel2021, Onida2002}
\begin{equation}
    \begin{pmatrix}
        \mathbf{A} & \mathbf{B} \\
       -\mathbf{B} & -\mathbf{A} 
    \end{pmatrix}
    \begin{pmatrix}
        X^s \\
        Y^s 
    \end{pmatrix} 
    =
    \begin{pmatrix}
        X^s \\
        Y^s 
    \end{pmatrix} \Omega_s, \label{eq:casida1}
\end{equation}
The dimension of matrices $\mathbf{A}$ and $\mathbf{B}$ is $N_{vc}^2$, where $N_{vc} = N_v \cdot N_c$ scales as $O(N^2)$ with respect to system size $N$.
Here $N_v$ and $N_c$ represent the number of occupied and empty orbitals, respectively.
With the random-phase approximation (RPA) used in the GW approximation, the matrix elements of $\mathbf{A}$ and $\mathbf{B}$ are given as
\begin{eqnarray}
    \mathbf{A}_{vc, v'c'} & = & (\epsilon_c-\epsilon_v)\delta_{vv'}\delta_{cc'} + (vc|v'c') \\
    \mathbf{B}_{vc, v'c'} & = & (vc|c'v')  \nonumber \\
    & = & \int \int \frac{ \phi_v (\mathbf{r}) \phi_c (\mathbf{r}) \phi_{c'}(\mathbf{r}') \phi_{v'}(\mathbf{r}') } { | \mathbf{r}-\mathbf{r}' | } d^3 r d^3 r' 
\end{eqnarray}
We use indices $v$ and $v'$ for occupied states, $c$ and $c'$ for empty states, and $k$, $l$, $n$, and $m$ for general orbitals, respectively. 
For finite systems, the Casida equation can be reformulated as a smaller eigenvalue problem~\cite{BRUNEVAL2016149,Tiago2006}
\begin{eqnarray}
    \mathbf{C} Z^s & = & Z^s \Omega^2_s \label{eq:casida2} 
\end{eqnarray}
where $\mathbf{C} =  (\mathbf{A}-\mathbf{B})^{1/2} (\mathbf{A}+\mathbf{B}) (\mathbf{A}-\mathbf{B})^{1/2}$ is a symmetric matrix of dimension $N^2_{vc}$. 
After solving Eq.~\ref{eq:casida2} for the eigenpairs $(Z^s, \Omega^2_s)$ of $\mathbf{C}$, one can compute the full-frequency GW self-energy 
\begin{eqnarray}
    \langle \phi_m | \Sigma^\mathbf{GW} (\omega) | \phi_m \rangle =  \Sigma^\mathrm{ex}_{mm} + \Sigma^\mathrm{corr}_{mm} (\omega), \\
    \Sigma^\mathrm{ex}_{mm} = - \sum_{v} (v m | v m ),  \\
    \Sigma^\mathrm{corr}_{mm} (\omega) = \sum^{N_v+N_c}_{n} \sum^{N_{vc}}_{s} \frac{W^s_{nm} W
^s_{nm}} { \omega - \epsilon_n + \eta_n (\Omega_s - i \delta )  } \label{eq:sigma_c}
\end{eqnarray}
where $\eta_n$ is 1 for occupied orbitals and -1 for empty orbitals, and $\delta$ is a positive infinitesimal number to avoid singularity. 
The matrix elements $W^s_{nm}$ are  
\begin{eqnarray}
    W^s_{nm} & = & \sum_{vc} (nm|vc) \sqrt{\frac{\epsilon_c - \epsilon_v}{\Omega_s}} Z^s_{vc} .
\end{eqnarray}

The exchange part of the self-energy $\Sigma^\mathrm{ex}_{mm}$ is independent of frequency and relatively easy to compute, while the correlation part $\Sigma^\mathbf{corr}_{mm}(\omega)$ includes the frequency-dependent screening effects of dielectric responses. 
The poles of frequency-dependent screened effects can be determined by the eigenvalues of $\mathbf{C}$. 
As a result, the most expensive step of the aforementioned method is diagonalizing the Casida equation, as the computational cost scales as $O(N^6)$. 
To make further progress, we intend to avoid this costly step by defining a vector $|P_{nm}\rangle$ of dimension $N_{vc}$, which has elements given by $(P_{nm})_{vc} = (nm|vc) (\epsilon_c - \epsilon_v)^{1/2}$. 
Then $W^s_{nm}$ becomes
\begin{eqnarray}
    W^s_{nm} & = & \langle P_{nm} | Z^s \rangle \Omega_s^{-\frac{1}{2}}. 
\end{eqnarray}
$\Sigma^\mathrm{corr}_{mm}$ can be rewritten as
\begin{eqnarray} 
    \Sigma^\mathrm{corr}_{mm} (\omega) & = & \sum^{N_v+N_c}_{n=1} \Sigma^\mathrm{corr}_{mm} (\omega, n), \label{eq:sigma_sum} 
\end{eqnarray}
where
\begin{eqnarray}
    \Sigma^\mathrm{corr}_{mm} (\omega, n) 
    & = & \frac{1}{z_n} \sum^{N_{vc}}_{s=1} \langle P_{nm} |  Z^s \rangle \langle Z^s | P_{nm} \rangle \times \nonumber \\
    & & \Big[ \frac{1}{\Omega_s} - \frac{1}{\Omega_s + \eta_n z_n } \Big], \label{eq:sigma_c2}
\end{eqnarray}
where $z_n = \omega - \epsilon_n - i \eta_n \delta $.

Examining Eq.~\ref{eq:sigma_c2}, we note the formula for $\Sigma^\mathrm{corr}_{mm}(\omega, n)$ is similar to a general resolvent matrix element of the form $\sum_{k} \langle \star | k \rangle \langle k | \star \rangle /(z-\lambda_k) = \langle \star | 1/(z - \mathbf{H}) | \star \rangle $,
where $\mathbf{H}$ is a general Hermitian matrix with eigenvalues $\lambda_k$ and eigenvectors $|k\rangle $, $z$ is a complex number, and $| \star \rangle $ is a ket. 
Motivated by this observation, we reformulate Eq.~\ref{eq:sigma_c2} as the resolvent of a symmetric matrix $\mathbf{D}$
\begin{eqnarray}
    \Sigma^\mathrm{corr}_{mm} (\omega, n) & = & \frac{1}{z_n}  \langle P_{nm} | \frac{1}{\mathbf{D}} - \frac{1}{\mathbf{D}+\eta_n z_n} | P_{nm} \rangle . \label{eq:resolvent}
\end{eqnarray}
Matrix $\mathbf{D}$ satisfies $\mathbf{D}^2=\mathbf{C}$ and its eigenvalues are the square root of those of matrix C, i.e.,  $\mathbf{D} Z^s= Z^s \Omega_s$.
We use a $g$-th degree polynomial function $p_g$ to fit the square root function $p_g (x) = \sum^{g}_{k=0} a_k x^k \approx \sqrt{x} $ within $x \in [ \min{\Omega^2_s}, \max{\Omega^2_s} ]$, which is the range between minimum and maximum eigenvalues of matrix $\mathbf{C}$. 
Accordingly, $\mathbf{D}$ can be approximated by $\mathbf{D} = p_g (\mathbf{C}) + \Delta_g \approx p_g(\mathbf{C}) = a_0 \mathbf{I} + \sum^{g}_{k=1} a_k \mathbf{C}^k$, where $\mathbf{I}$ is an identity matrix and the fitting error $\Delta_g$ can be controlled via the degree $g$ of the polynomial function and fitting procedures. More discussions on Eq.~\ref{eq:sigma_sum} to Eq.~\ref{eq:resolvent} are presented in Section 1 of the Supplemental Material~\cite{supp}.

Given Eq.~\ref{eq:resolvent} and matrix $\mathbf{D}$, the Lanczos method can then be applied to efficiently compute the resolvent of matrix $\mathbf{D}$, which is an important step in calculating $\Sigma^\mathrm{corr}_{mm} (\omega, n)$. 
In the calculation of $\Sigma^\mathrm{corr}_{mm} (\omega)$, we prepare $|P_{nm}\rangle$ for each state $n$ in the summation of Eq.~\ref{eq:sigma_sum}, where
$|P_{nm}\rangle$ is used as the starting vector for the Lanczos tri-diagonalization procedure of the symmetric matrix $\mathbf{D}$. 
With $L$ steps of Lanczos iterations, one can construct a tridiagonal matrix $\mathbf{D}_L$ with dimension $L$ in the following form:
\begin{eqnarray}
    \mathbf{D}_L = 
    \begin{pmatrix}
    a_0 & b_1 & 0   & \ldots & 0 & 0 \\
    b_1 & a_1 & b_2 & \ldots & 0 & 0 \\
    0   & b_2 & a_2 & \ldots & 0 & 0 \\
    \vdots & \vdots & \vdots & \ddots & \vdots & \vdots \\
    0   &  0  &  0  & \ldots & a_{L-1} & b_L \\
    0   &  0  &  0  & \ldots & b_L & a_L 
    \end{pmatrix}.
\end{eqnarray}

Once the tridiagonal matrix $\mathbf{D}_L$ is obtained, a resolvent matrix element (such as  Eq.~\ref{eq:resolvent}) can be computed using the continuous fraction
\begin{eqnarray}
    \langle P_{nm} | \frac{1}{z-\mathbf{D}} | P_{nm} \rangle = \frac{1}{z-a_0-\frac{b^2_1}{z-a_1-\frac{b^2_2}{a_2-\ldots}}}, \label{eq:haydock}
\end{eqnarray}
which is also known as the Haydock method~\cite{HAYDOCK198011}. 
The computation of Eq.~\ref{eq:haydock} is efficient, and one can easily calculate the quasiparticle energies for a series of frequencies by varying $z$ in Eq.~\ref{eq:haydock}. 
When applied to eigenvalue problems, the Lanczos algorithm can lead to ghost eigenvalues. However, applying the Lanczos method to calculate resolvent is free of such a numerical problem~\cite{meyer1989}.

There are several advantages to using the Lanczos method for computing $\Sigma^\mathrm{corr}_{mm}$. 
Solving the eigenvalue problem of the Casida matrix $\mathbf{C}$ is avoided, and the resulting full-frequency GW calculations become more efficient than the conventional method represented by Eq.~\ref{eq:sigma_c}, which explicitly requires the eigenpairs of $\mathbf{C}$. 
Frequency grids, analytical continuation, and approximations like plasmon-pole models are not required, as the frequency dependence of $W$ and $\Sigma^\mathrm{corr}_{mm}$ are implicitly treated via Lanczos iterations. 
Moreover, the method is in principle applicable to any basis sets of wave functions, as our derivation does not rely on any features of specific basis functions. 

As a general-purpose algorithm, Lanczos-based methods have been used in computational material science, such as computing Green's function~\cite{HAYDOCK198011}, optical absorption spectra with linear-response time-dependent density functional theory~\cite{MALCIOGLU20111744, walker2006, zamok2022, GRUNING20112148, Benedict1998} and Bethe-Salpeter equation~\cite{shao2018}. 
Earlier work~\cite{govoni2015, Umari2010, laflamme2015} applied Lanczos methods for solving the Sternheimer equation to obtain frequency-dependent screened Coulomb potential. 
Recently, several new methods~\cite{bintrim2021, scott2023,backhouse2021,chiarotti2022,leon2023} have been explored to achieve efficient full-frequency GW calculations.
For example, Scott et al.~\cite{scott2023} adopted a block Lanczos algorithm to solve an effective Hamiltonian whose eigenvalues systematically approximate the excitation energies of GW theory. 
Bintrim and Berkelbach~\cite{bintrim2021,bintrim2022} proposed a method that does not require integration over the frequency grids. Instead, the GW quasiparticle energies are obtained by solving the eigenvalues of an effective Hamiltonian, which follows the algebraic diagrammatic construction~\cite{schirmer1983}.
Compared to these methods, our method does not solve the Sternheimer equation or obtain GW quasiparticle energies from the eigenvalues of an effective Hamiltonian. 
Instead, the frequency-dependent screened Coulomb potential is found using linear-response TDDFT within the Casida formalism, and the GW quasiparticle energies are computed from a summation of resolvent elements given by Eq.~\ref{eq:resolvent}. 

\begin{figure}[htb!]
    \centering
    \includegraphics[width=0.32\textwidth]{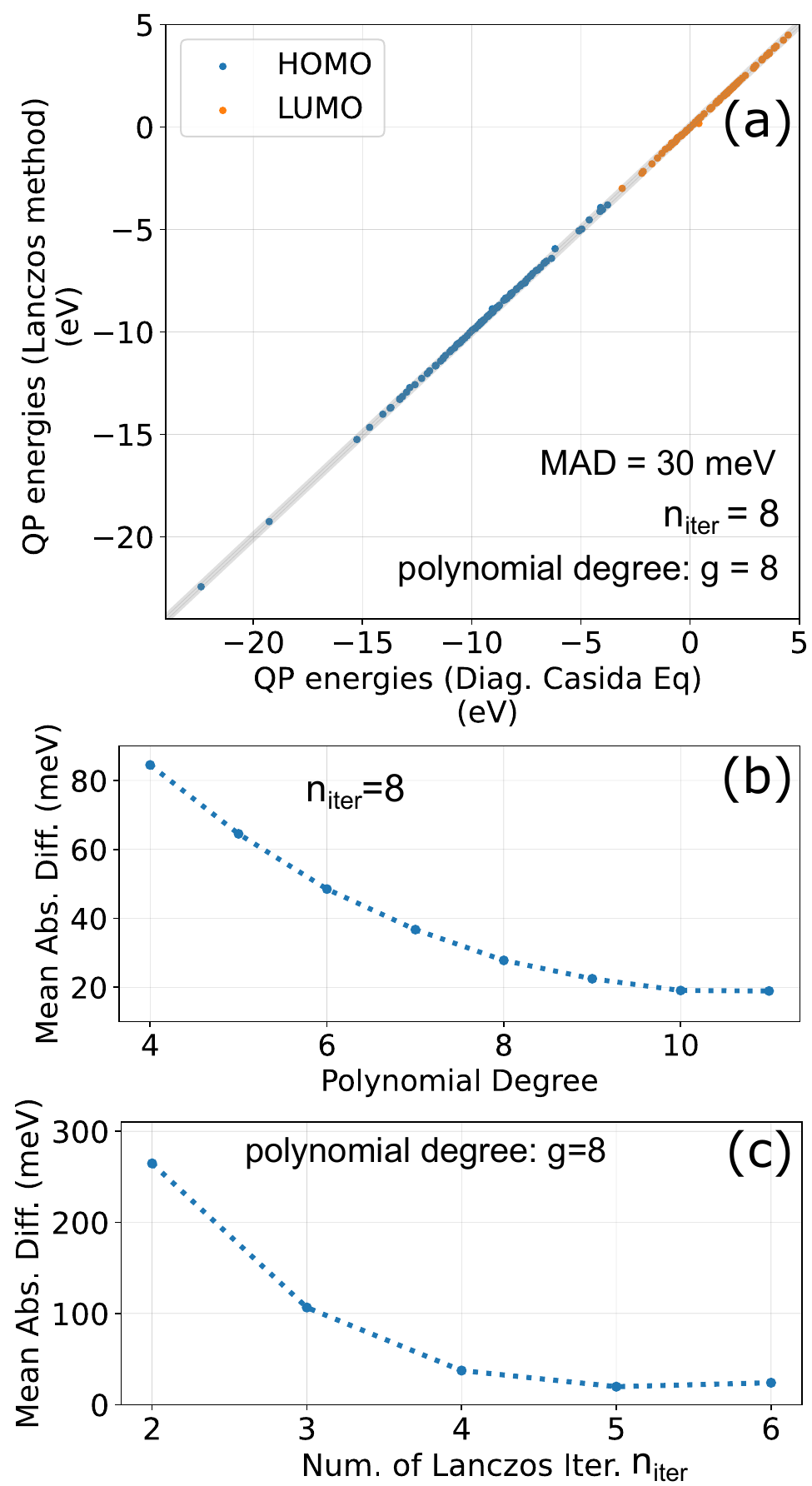}
    \caption{ (a). Comparison between the HOMO and LUMO energies calculated with the reference method and the Lanczos method.
    (b). The mean absolute differences (MAD) between the quasiparticle energies of one hundred small molecules calculated with the reference approach and the Lanczos method with different degrees of polynomial functions; 
    (c). MAD of quasiparticle energies calculated with the reference method and the Lanczos method using different numbers of iterations $N_\mathrm{iter}$; 
     }
    \label{fig:gw100}
\end{figure}

The accuracy of the Lanczos-based method for the $GW$ approximation is checked by calculating the highest occupied and lowest unoccupied molecule orbital (HOMO/LUMO) energies of the GW100 set~\cite{vansetten2015,govoni2018,forster2021,gao2019_gw100}, which include 100 small close-shell molecules for benchmarking different implementations of the GW approximation. $G_0 W_0$-level calculations are carried out throughout this work.
As studied in previous work, $G_0 W_0$-level calculations depend on the starting point, while quasiparticle self-consistent GW (QSGW) and fully self-consistent GW can alleviate the dependence of calculation results on the starting points~\cite{caruso2013,marom2012,Rinke_2005,Dauth2016}. Our new Lanczos method is compatible with QSGW~\cite{Hung2016} because the accelerated steps (i.e., bypassing the diagonalization of the Casida equation and using the Lanczos method to compute the correlation part of the self-energy) do not interfere with the self-consistent iterations. The Lanczos method only requires the updated quasiparticle energies and wave functions of the current iteration to start the next iteration of GW calculation. 
A real-space-based pseudo-potential DFT code \texttt{PARSEC} is used in our implementation to efficiently obtain Kohn-Sham orbitals for large finite systems~\cite{Kronik2006,dogan2023}. 
More details of our computations are presented in Section 2 of the Supplemental Material~\cite{supp,perdew1996,troullier1991}.
Fig.~\ref{fig:gw100} (a) shows the results computed with the Lanczos method and the reference agrees well for all GW100 molecules. The mean average difference (MAD) between the results calculated using the reference method, which finds the eigenpairs of the Casida equation explicitly, and the Lanczos method is within 20 meV. 
Our tests also show the Lanczos-based formalism converges fast to the degree $g$ of polynomial $p_g$ and the number of Lanczos iterations $N_\textrm{iter}$. 
As shown in Fig.~\ref{fig:gw100} (b) and (c), the MAD is below 30 meV when the polynomial degree $g \ge 8$ and $N_\textrm{iter} \ge 5$. 

The computationally expensive steps in our method are: (1) calculating electron-repulsion integrals $(kl|nm)$ and (2) calculating the matrix-vector product $ \mathbf{D} | \star \rangle $, where $| \star \rangle$ is a general vector. 
One can use suitable low-rank approximation methods, such as resolution-of-identity or density-fitting methods~\cite{weigend2002,Ren_2012,hu2020_ISDF_LRTDDFT}, to speed up these computations. 
Density-fitting methods exploit the rank deficiency of orbital pair products $\phi_n(\mathbf{r}) \phi_m(\mathbf{r})$ and use a set of auxiliary basis functions $\zeta_\mu (\mathbf{r}) $ to fit these orbital pairs
\begin{eqnarray}
    \phi_n(\mathbf{r}) \phi_m(\mathbf{r}) \approx \sum_{\mu=1}^{N_\mu} \zeta_\mu (\mathbf{r}) \mathcal{C}^\mu_{nm} \label{eq:df}
\end{eqnarray}
where the required number of auxiliary basis functions $N_\mu$ for accurately representing the orbital pairs is expected to be small and scale as $O(N)$ and $\mathcal{C}^\mu_{nm}$ are fitting coefficients. 
With the approximation given in Eq.~\ref{eq:df}, one can calculate integrals $(kl|nm)$, which contribute to the elements of $\mathbf{C}$ and $\mathbf{D}$, with the following equations
\begin{eqnarray}
    (kl|nm) \approx \sum_{\alpha=1}^{N_\mu} \sum_{\beta=1}^{N_\mu} (\alpha|\beta) \mathcal{C}^\alpha_{kl} \mathcal{C}^\beta_{nm} \\
    (\alpha|\beta) = \int \frac{\zeta_\alpha(\mathbf{r}) \zeta_\beta(\mathbf{r}')}{|\mathbf{r}-\mathbf{r}'|} d^3r d^3r' .
\end{eqnarray}
These methods reduce four-center integrals to two-center integrals and also factorize $\mathbf{C}$ and $\mathbf{D}$ as products of small matrices to accelerate the matrix-vector products. 
Here we used the interpolative separable density fitting (ISDF) method to efficiently construct low-rank approximations of orbital pairs~\cite{LU2015329,Gao_2022,hu2020_ISDF_LRTDDFT}. 
Additionally, one can further exploit the point-group symmetries to make the Casida matrix $\mathbf{C}$ block diagonal and simplify the calculation of matrix-vector products~\cite{gao2020}. 

\begin{figure}[tb!]
    \centering
    \includegraphics[width=0.9\linewidth]{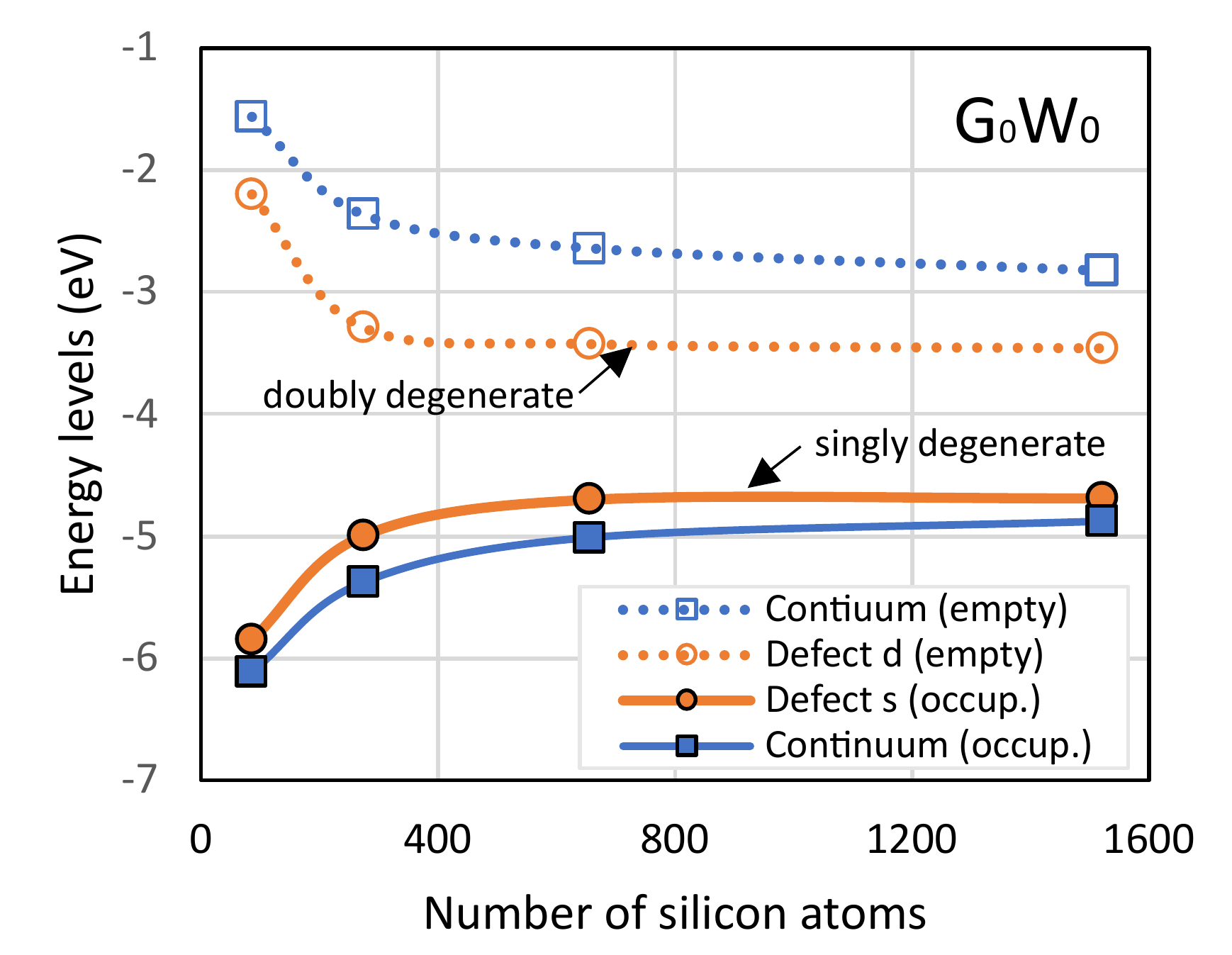}
    \caption{Evolution of ``continuum" states and silicon-vacancy defect states in different-sized silicon clusters. }
    \label{fig:Si_vacancy_defect}
\end{figure}

Combined with the ISDF method~\cite{Gao_2022}, our method is efficient and enables large-scale $G_0W_0$ computations with modest computing resources. 
To demonstrate the efficiency of our method, we performed calculations for hydrogen-passivated silicon clusters. 
Silicon clusters have attracted research interest as prototypical semiconducting clusters for studying the fundamental physical properties of zero-dimensional systems~\cite{Barbagiovanni2014} and their applications in many fields~\cite{liang2019, Dohnalová2013, neiner2006, Chang_2014}. 
Defects in passivated silicon nanocrystals can introduce mid-gap defect levels as potential sources of photoluminescence~\cite{Godefroo2008,Priolo2014}, while their electronic structures are rarely studied by GW calculations.
We computed the defect energy levels of charge-neutral silicon vacancies in silicon clusters of different sizes. 
The ground state of a silicon vacancy has zero net spin. 
Different from nano-diamondoids, where surface states are located in the gap, the surface states of silicon nanoclusters are mixed with continuum states and the mid-gap states originate from defects. 
In the single-particle level, an occupied singlet and a pair of unoccupied doubly degenerate defect states are located inside the gap~\cite{Liao2022}. 
As shown in Fig.~\ref{fig:Si_vacancy_defect}, when the size of silicon clusters increases, the energies of defect states evolve at a similar rate as the continuum states. 
For the Si$_{1522}$H$_{524}$ cluster, the band gap of continuum states is around 2.3 eV, still far from the bulk silicon band gap of 1.1~eV. 
We also computed the HOMO-LUMO gap of non-defective silicon nanocrystals, and our results agree well with previous calculations~\cite{govoni2015, Neuhauser2014} (see Section 3 in the Supplemental Material~\cite{supp} for more details).

\begin{table}
    \centering
    \begin{tabular}{c|ccc|cc}
    \hline \hline
        Nanocluster & $N_v$ & $N_c+N_v$ & $N_\mu$ & $N_\mathbf{node}$ & $t_\mathrm{wall}$ (hr) \\ 
       \hline
        Si$_{86}$H$_{76}$    & 210  & 1730   & 7000   & 1  & 0.02 \\
        Si$_{274}$H$_{172}$  & 634  & 5200   & 20000  & 2  & 0.3 \\
        Si$_{452}$H$_{228}$  & 1018 & 8200   & 32000  & 2  & 0.7 \\
        Si$_{656}$H$_{300}$  & 1462 & 12000  & 48000  & 2  & 1.5 \\
        Si$_{1522}$H$_{524}$ & 3306 & 27000  & 108000 & 10 & 8.5 \\ 
        Si$_{1947}$H$_{604}$ & 4196 & 33400  & 133600 & 20 & 9.2 \\
       \hline \hline
    \end{tabular}
    \caption{ The running time $t_\mathrm{wall}$ and the number of compute nodes $N_\mathrm{node}$ for calculating the GW quasiparticle energy of one quasiparticle state. 
    Each compute node has 64 cores and 4 graphic processing units (GPU). 
    $t_\mathrm{wall}$ includes the running time for performing the ISDF method and computing $\Sigma^\mathbf{GW}(\omega)$ using the Lanczos method. 
    $N_\mathbf{iter} = 8$ and $g = 7$ are used for Lanczos iterations and polynomial functions, respectively. }
    \label{tab:runningtime}
\end{table}

\begin{figure}[htb!]
    \centering
    \includegraphics[width=0.81\linewidth]{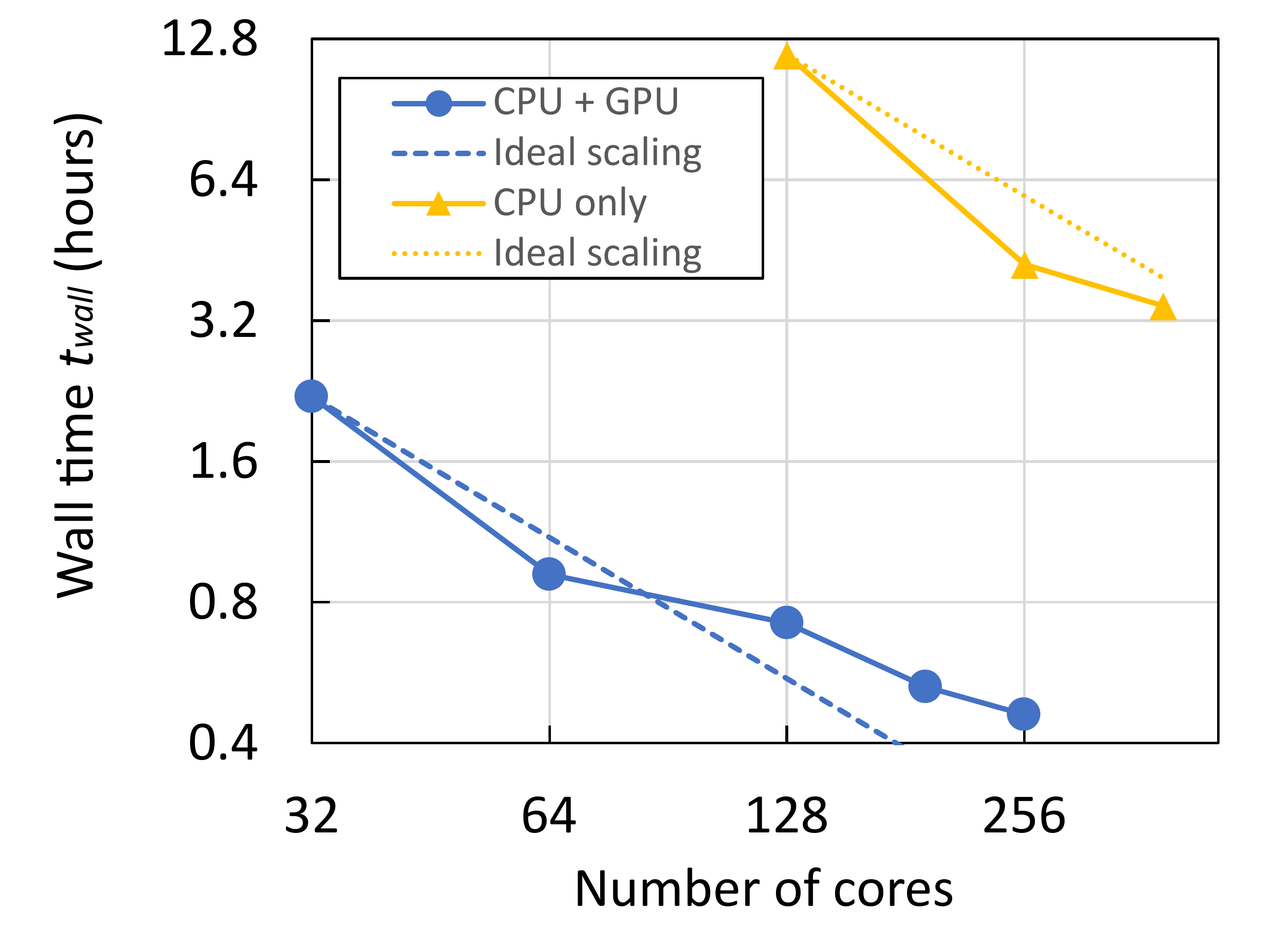}
    \caption{The wall time for computing the GW quasiparticle energy of one state of Si$_{452}$H$_{228}$ cluster with different numbers of CPU cores and GPUs (Nvidia A100). Every computer node has an EPYC 7763 CPU with 64 cores. For calculations accelerated with GPUs, every 16 CPU cores share 1 GPU processor. }
    \label{fig:Si453H228_scaling}
\end{figure}


The main calculation parameters and required computation resources for these calculations are shown in Table~\ref{tab:runningtime}. 
Notably, only two computing nodes are required for Si$_{656}$H$_{300}$.
For the largest system, Si$_{1953}$H$_{604}$ with a diameter of around 4~nm, we used 20 nodes to accomplish the full-frequency GW calculation.
The computational cost of our algorithm has a theoretical scaling of $O(N^4)$. 
In the benchmarks of silicon clusters, we observe a practical scaling of roughly $O(N^{2.3})$ for systems with less than 600 silicon atoms (see Section 4 of the Supplemental Material~\cite{supp} for a detailed analysis of the computational costs). 
As shown in Fig.~\ref{fig:Si453H228_scaling}, we compared the running time for full-frequency GW calculations of Si$_{453}$H$_{228}$ using different numbers of nodes.
For our Lanczos-based method, the most time-consuming steps are matrix-matrix and matrix-vector multiplications, which are suitable for massive parallelization and acceleration with GPUs in heterogeneous supercomputers. Fig.~\ref{fig:Si453H228_scaling} demonstrates the reasonably good strong scaling with computation resources. 
When the calculations are accelerated with GPUs, the speed-up factors compared to CPU-only calculations are around 20. 


In summary, a full-frequency GW formalism based on a Lanczos method is proposed to realize efficient modeling of hundreds of atoms with modest resources.
This method can be used for highly efficient full-frequency GW calculations of large finite systems, such as semiconductor quantum dots and ligand-protected superatomic clusters with a few hundred atoms. 
Our method can also facilitate the construction of computational databases with quasiparticle-energy data, which were challenging to accomplish with limited computational costs before. 
This method is ready to generalize to extended systems, for which complex-valued wave functions are required. 
If the Tamm-Dancoff approximation is used (i.e., setting matrix $\mathbf{B}=0$ in the RPA Casida matrix) for extended systems, then the calculation is greatly simplified as only a Hermitian matrix $\mathbf{A}$ remains, and a standard Lanczos algorithm for Hermitian matrix can be used. 
On the other hand, if RPA is used, then a Lanczos-based method designed for pseudo-Hermitian matrices~\cite{GRUNING20112148, brabec2015, shao2018} can be adopted. 

We are grateful for the discussion with Peihong Zhang. This work is supported by the National Natural Science Foundation of China (12104080, 91961204), GHfund A (2022201, ghfund202202012538), the Fundamental Research Funds for the Central Universities (DUT22LK04 and DUT22ZD103), and a sub-award
from the Center for Computational Study of Excited-State Phenomena in Energy Materials at the Lawrence Berkeley National Laboratory, which is funded by the U.S. Department of Energy, Office of Science, Basic Energy Sciences, Materials
Sciences and Engineering Division under Contract No. DEAC02-05CH11231, as part of the Computational Materials Sciences Program. 
Computational resources are provided by the Perlmutter supercomputer cluster of the National Energy Research Scientific Computing Center (NERSC), the Sugon Supercomputer Center at Wuzhen, and the Sugon Supercomputer Center at Kunshan.

\bibliography{GW_lanczos}

\begin{thebibliography}{74}%
\makeatletter
\providecommand \@ifxundefined [1]{%
 \@ifx{#1\undefined}
}%
\providecommand \@ifnum [1]{%
 \ifnum #1\expandafter \@firstoftwo
 \else \expandafter \@secondoftwo
 \fi
}%
\providecommand \@ifx [1]{%
 \ifx #1\expandafter \@firstoftwo
 \else \expandafter \@secondoftwo
 \fi
}%
\providecommand \natexlab [1]{#1}%
\providecommand \enquote  [1]{``#1''}%
\providecommand \bibnamefont  [1]{#1}%
\providecommand \bibfnamefont [1]{#1}%
\providecommand \citenamefont [1]{#1}%
\providecommand \href@noop [0]{\@secondoftwo}%
\providecommand \href [0]{\begingroup \@sanitize@url \@href}%
\providecommand \@href[1]{\@@startlink{#1}\@@href}%
\providecommand \@@href[1]{\endgroup#1\@@endlink}%
\providecommand \@sanitize@url [0]{\catcode `\\12\catcode `\$12\catcode
  `\&12\catcode `\#12\catcode `\^12\catcode `\_12\catcode `\%12\relax}%
\providecommand \@@startlink[1]{}%
\providecommand \@@endlink[0]{}%
\providecommand \url  [0]{\begingroup\@sanitize@url \@url }%
\providecommand \@url [1]{\endgroup\@href {#1}{\urlprefix }}%
\providecommand \urlprefix  [0]{URL }%
\providecommand \Eprint [0]{\href }%
\providecommand \doibase [0]{https://doi.org/}%
\providecommand \selectlanguage [0]{\@gobble}%
\providecommand \bibinfo  [0]{\@secondoftwo}%
\providecommand \bibfield  [0]{\@secondoftwo}%
\providecommand \translation [1]{[#1]}%
\providecommand \BibitemOpen [0]{}%
\providecommand \bibitemStop [0]{}%
\providecommand \bibitemNoStop [0]{.\EOS\space}%
\providecommand \EOS [0]{\spacefactor3000\relax}%
\providecommand \BibitemShut  [1]{\csname bibitem#1\endcsname}%
\let\auto@bib@innerbib\@empty
\bibitem [{\citenamefont {Hedin}(1965)}]{hedin1965}%
  \BibitemOpen
  \bibfield  {author} {\bibinfo {author} {\bibfnamefont {L.}~\bibnamefont
  {Hedin}},\ }\bibfield  {title} {\bibinfo {title} {New method for calculating
  the one-particle green's function with application to the electron-gas
  problem},\ }\href {https://doi.org/10.1103/PhysRev.139.A796} {\bibfield
  {journal} {\bibinfo  {journal} {Phys. Rev.}\ }\textbf {\bibinfo {volume}
  {139}},\ \bibinfo {pages} {A796} (\bibinfo {year} {1965})}\BibitemShut
  {NoStop}%
\bibitem [{\citenamefont {Hybertsen}\ and\ \citenamefont
  {Louie}(1986)}]{hybertsen1986}%
  \BibitemOpen
  \bibfield  {author} {\bibinfo {author} {\bibfnamefont {M.~S.}\ \bibnamefont
  {Hybertsen}}\ and\ \bibinfo {author} {\bibfnamefont {S.~G.}\ \bibnamefont
  {Louie}},\ }\bibfield  {title} {\bibinfo {title} {Electron correlation in
  semiconductors and insulators: Band gaps and quasiparticle energies},\ }\href
  {https://doi.org/10.1103/PhysRevB.34.5390} {\bibfield  {journal} {\bibinfo
  {journal} {Phys. Rev. B}\ }\textbf {\bibinfo {volume} {34}},\ \bibinfo
  {pages} {5390} (\bibinfo {year} {1986})}\BibitemShut {NoStop}%
\bibitem [{\citenamefont {Godby}\ \emph {et~al.}(1988)\citenamefont {Godby},
  \citenamefont {Schl\"uter},\ and\ \citenamefont {Sham}}]{godby1988}%
  \BibitemOpen
  \bibfield  {author} {\bibinfo {author} {\bibfnamefont {R.~W.}\ \bibnamefont
  {Godby}}, \bibinfo {author} {\bibfnamefont {M.}~\bibnamefont {Schl\"uter}},\
  and\ \bibinfo {author} {\bibfnamefont {L.~J.}\ \bibnamefont {Sham}},\
  }\bibfield  {title} {\bibinfo {title} {Self-energy operators and
  exchange-correlation potentials in semiconductors},\ }\href
  {https://doi.org/10.1103/PhysRevB.37.10159} {\bibfield  {journal} {\bibinfo
  {journal} {Phys. Rev. B}\ }\textbf {\bibinfo {volume} {37}},\ \bibinfo
  {pages} {10159} (\bibinfo {year} {1988})}\BibitemShut {NoStop}%
\bibitem [{\citenamefont {Rohlfing}\ and\ \citenamefont
  {Louie}(1998)}]{rohlfing1998}%
  \BibitemOpen
  \bibfield  {author} {\bibinfo {author} {\bibfnamefont {M.}~\bibnamefont
  {Rohlfing}}\ and\ \bibinfo {author} {\bibfnamefont {S.~G.}\ \bibnamefont
  {Louie}},\ }\bibfield  {title} {\bibinfo {title} {Electron-hole excitations
  in semiconductors and insulators},\ }\href
  {https://doi.org/10.1103/PhysRevLett.81.2312} {\bibfield  {journal} {\bibinfo
   {journal} {Phys. Rev. Lett.}\ }\textbf {\bibinfo {volume} {81}},\ \bibinfo
  {pages} {2312} (\bibinfo {year} {1998})}\BibitemShut {NoStop}%
\bibitem [{\citenamefont {Albrecht}\ \emph {et~al.}(1998)\citenamefont
  {Albrecht}, \citenamefont {Reining}, \citenamefont {Del~Sole},\ and\
  \citenamefont {Onida}}]{Albrecht1998}%
  \BibitemOpen
  \bibfield  {author} {\bibinfo {author} {\bibfnamefont {S.}~\bibnamefont
  {Albrecht}}, \bibinfo {author} {\bibfnamefont {L.}~\bibnamefont {Reining}},
  \bibinfo {author} {\bibfnamefont {R.}~\bibnamefont {Del~Sole}},\ and\
  \bibinfo {author} {\bibfnamefont {G.}~\bibnamefont {Onida}},\ }\bibfield
  {title} {\bibinfo {title} {Ab initio calculation of excitonic effects in the
  optical spectra of semiconductors},\ }\href
  {https://doi.org/10.1103/PhysRevLett.80.4510} {\bibfield  {journal} {\bibinfo
   {journal} {Phys. Rev. Lett.}\ }\textbf {\bibinfo {volume} {80}},\ \bibinfo
  {pages} {4510} (\bibinfo {year} {1998})}\BibitemShut {NoStop}%
\bibitem [{\citenamefont {Chan}\ \emph {et~al.}(2021)\citenamefont {Chan},
  \citenamefont {Qiu}, \citenamefont {da~Jornada},\ and\ \citenamefont
  {Louie}}]{yanghao2021}%
  \BibitemOpen
  \bibfield  {author} {\bibinfo {author} {\bibfnamefont {Y.-H.}\ \bibnamefont
  {Chan}}, \bibinfo {author} {\bibfnamefont {D.~Y.}\ \bibnamefont {Qiu}},
  \bibinfo {author} {\bibfnamefont {F.~H.}\ \bibnamefont {da~Jornada}},\ and\
  \bibinfo {author} {\bibfnamefont {S.~G.}\ \bibnamefont {Louie}},\ }\bibfield
  {title} {\bibinfo {title} {Giant exciton-enhanced shift currents and direct
  current conduction with subbandgap photo excitations produced by
  many-electron interactions},\ }\href
  {https://doi.org/10.1073/pnas.1906938118} {\bibfield  {journal} {\bibinfo
  {journal} {Proceedings of the National Academy of Sciences}\ }\textbf
  {\bibinfo {volume} {118}},\ \bibinfo {pages} {e1906938118} (\bibinfo {year}
  {2021})}\BibitemShut {NoStop}%
\bibitem [{\citenamefont {Biermann}\ \emph {et~al.}(2003)\citenamefont
  {Biermann}, \citenamefont {Aryasetiawan},\ and\ \citenamefont
  {Georges}}]{Biermann2003}%
  \BibitemOpen
  \bibfield  {author} {\bibinfo {author} {\bibfnamefont {S.}~\bibnamefont
  {Biermann}}, \bibinfo {author} {\bibfnamefont {F.}~\bibnamefont
  {Aryasetiawan}},\ and\ \bibinfo {author} {\bibfnamefont {A.}~\bibnamefont
  {Georges}},\ }\bibfield  {title} {\bibinfo {title} {First-principles approach
  to the electronic structure of strongly correlated systems: Combining the
  ${GW}$ approximation and dynamical mean-field theory},\ }\href
  {https://doi.org/10.1103/PhysRevLett.90.086402} {\bibfield  {journal}
  {\bibinfo  {journal} {Phys. Rev. Lett.}\ }\textbf {\bibinfo {volume} {90}},\
  \bibinfo {pages} {086402} (\bibinfo {year} {2003})}\BibitemShut {NoStop}%
\bibitem [{\citenamefont {Sun}\ and\ \citenamefont {Kotliar}(2002)}]{Ping2002}%
  \BibitemOpen
  \bibfield  {author} {\bibinfo {author} {\bibfnamefont {P.}~\bibnamefont
  {Sun}}\ and\ \bibinfo {author} {\bibfnamefont {G.}~\bibnamefont {Kotliar}},\
  }\bibfield  {title} {\bibinfo {title} {Extended dynamical mean-field theory
  and $\mathrm{GW}$ method},\ }\href
  {https://doi.org/10.1103/PhysRevB.66.085120} {\bibfield  {journal} {\bibinfo
  {journal} {Phys. Rev. B}\ }\textbf {\bibinfo {volume} {66}},\ \bibinfo
  {pages} {085120} (\bibinfo {year} {2002})}\BibitemShut {NoStop}%
\bibitem [{\citenamefont {Zhu}\ and\ \citenamefont {Chan}(2021)}]{Tianyu2021}%
  \BibitemOpen
  \bibfield  {author} {\bibinfo {author} {\bibfnamefont {T.}~\bibnamefont
  {Zhu}}\ and\ \bibinfo {author} {\bibfnamefont {G.~K.-L.}\ \bibnamefont
  {Chan}},\ }\bibfield  {title} {\bibinfo {title} {Ab initio full cell
  ${GW}+\mathrm{DMFT}$ for correlated materials},\ }\href
  {https://doi.org/10.1103/PhysRevX.11.021006} {\bibfield  {journal} {\bibinfo
  {journal} {Phys. Rev. X}\ }\textbf {\bibinfo {volume} {11}},\ \bibinfo
  {pages} {021006} (\bibinfo {year} {2021})}\BibitemShut {NoStop}%
\bibitem [{\citenamefont {Li}\ \emph {et~al.}(2019)\citenamefont {Li},
  \citenamefont {Antonius}, \citenamefont {Wu}, \citenamefont {da~Jornada},\
  and\ \citenamefont {Louie}}]{zhenglu2019}%
  \BibitemOpen
  \bibfield  {author} {\bibinfo {author} {\bibfnamefont {Z.}~\bibnamefont
  {Li}}, \bibinfo {author} {\bibfnamefont {G.}~\bibnamefont {Antonius}},
  \bibinfo {author} {\bibfnamefont {M.}~\bibnamefont {Wu}}, \bibinfo {author}
  {\bibfnamefont {F.~H.}\ \bibnamefont {da~Jornada}},\ and\ \bibinfo {author}
  {\bibfnamefont {S.~G.}\ \bibnamefont {Louie}},\ }\bibfield  {title} {\bibinfo
  {title} {Electron-phonon coupling from ab initio linear-response theory
  within the $gw$ method: Correlation-enhanced interactions and
  superconductivity in {Ba}$_{1\ensuremath{-}x}${K}$_{x}${BiO}$_{3}$},\ }\href
  {https://doi.org/10.1103/PhysRevLett.122.186402} {\bibfield  {journal}
  {\bibinfo  {journal} {Phys. Rev. Lett.}\ }\textbf {\bibinfo {volume} {122}},\
  \bibinfo {pages} {186402} (\bibinfo {year} {2019})}\BibitemShut {NoStop}%
\bibitem [{\citenamefont {Li}\ \emph {et~al.}(2021)\citenamefont {Li},
  \citenamefont {Wu}, \citenamefont {Chan},\ and\ \citenamefont
  {Louie}}]{zhenglu2021}%
  \BibitemOpen
  \bibfield  {author} {\bibinfo {author} {\bibfnamefont {Z.}~\bibnamefont
  {Li}}, \bibinfo {author} {\bibfnamefont {M.}~\bibnamefont {Wu}}, \bibinfo
  {author} {\bibfnamefont {Y.-H.}\ \bibnamefont {Chan}},\ and\ \bibinfo
  {author} {\bibfnamefont {S.~G.}\ \bibnamefont {Louie}},\ }\bibfield  {title}
  {\bibinfo {title} {Unmasking the origin of kinks in the photoemission spectra
  of cuprate superconductors},\ }\href
  {https://doi.org/10.1103/PhysRevLett.126.146401} {\bibfield  {journal}
  {\bibinfo  {journal} {Phys. Rev. Lett.}\ }\textbf {\bibinfo {volume} {126}},\
  \bibinfo {pages} {146401} (\bibinfo {year} {2021})}\BibitemShut {NoStop}%
\bibitem [{\citenamefont {Golze}\ \emph {et~al.}(2019)\citenamefont {Golze},
  \citenamefont {Dvorak},\ and\ \citenamefont {Rinke}}]{golze2019}%
  \BibitemOpen
  \bibfield  {author} {\bibinfo {author} {\bibfnamefont {D.}~\bibnamefont
  {Golze}}, \bibinfo {author} {\bibfnamefont {M.}~\bibnamefont {Dvorak}},\ and\
  \bibinfo {author} {\bibfnamefont {P.}~\bibnamefont {Rinke}},\ }\bibfield
  {title} {\bibinfo {title} {The {GW} compendium: A practical guide to
  theoretical photoemission spectroscopy},\ }\href
  {https://doi.org/10.3389/fchem.2019.00377} {\bibfield  {journal} {\bibinfo
  {journal} {Frontiers in Chemistry}\ }\textbf {\bibinfo {volume} {7}},\
  \bibinfo {pages} {377} (\bibinfo {year} {2019})}\BibitemShut {NoStop}%
\bibitem [{\citenamefont {Gao}\ \emph {et~al.}(2022)\citenamefont {Gao},
  \citenamefont {Xia}, \citenamefont {Zhang}, \citenamefont {Chelikowsky},\
  and\ \citenamefont {Zhao}}]{Gao_2022}%
  \BibitemOpen
  \bibfield  {author} {\bibinfo {author} {\bibfnamefont {W.}~\bibnamefont
  {Gao}}, \bibinfo {author} {\bibfnamefont {W.}~\bibnamefont {Xia}}, \bibinfo
  {author} {\bibfnamefont {P.}~\bibnamefont {Zhang}}, \bibinfo {author}
  {\bibfnamefont {J.~R.}\ \bibnamefont {Chelikowsky}},\ and\ \bibinfo {author}
  {\bibfnamefont {J.}~\bibnamefont {Zhao}},\ }\bibfield  {title} {\bibinfo
  {title} {Numerical methods for efficient {GW} calculations and the
  applications in low-dimensional systems},\ }\href
  {https://doi.org/10.1088/2516-1075/ac709a} {\bibfield  {journal} {\bibinfo
  {journal} {Electronic Structure}\ }\textbf {\bibinfo {volume} {4}},\ \bibinfo
  {pages} {023003} (\bibinfo {year} {2022})}\BibitemShut {NoStop}%
\bibitem [{\citenamefont {Wilhelm}\ \emph {et~al.}(2018)\citenamefont
  {Wilhelm}, \citenamefont {Golze}, \citenamefont {Talirz}, \citenamefont
  {Hutter},\ and\ \citenamefont {Pignedoli}}]{wilhelm2018}%
  \BibitemOpen
  \bibfield  {author} {\bibinfo {author} {\bibfnamefont {J.}~\bibnamefont
  {Wilhelm}}, \bibinfo {author} {\bibfnamefont {D.}~\bibnamefont {Golze}},
  \bibinfo {author} {\bibfnamefont {L.}~\bibnamefont {Talirz}}, \bibinfo
  {author} {\bibfnamefont {J.}~\bibnamefont {Hutter}},\ and\ \bibinfo {author}
  {\bibfnamefont {C.~A.}\ \bibnamefont {Pignedoli}},\ }\bibfield  {title}
  {\bibinfo {title} {Toward {GW} calculations on thousands of atoms},\ }\href
  {https://doi.org/10.1021/acs.jpclett.7b02740} {\bibfield  {journal} {\bibinfo
   {journal} {The Journal of Physical Chemistry Letters}\ }\textbf {\bibinfo
  {volume} {9}},\ \bibinfo {pages} {306} (\bibinfo {year} {2018})}\BibitemShut
  {NoStop}%
\bibitem [{\citenamefont {Vlček}\ \emph {et~al.}(2018)\citenamefont {Vlček},
  \citenamefont {Li}, \citenamefont {Baer}, \citenamefont {Rabani},\ and\
  \citenamefont {Neuhauser}}]{vlcek2018}%
  \BibitemOpen
  \bibfield  {author} {\bibinfo {author} {\bibfnamefont {V.}~\bibnamefont
  {Vlček}}, \bibinfo {author} {\bibfnamefont {W.}~\bibnamefont {Li}}, \bibinfo
  {author} {\bibfnamefont {R.}~\bibnamefont {Baer}}, \bibinfo {author}
  {\bibfnamefont {E.}~\bibnamefont {Rabani}},\ and\ \bibinfo {author}
  {\bibfnamefont {D.}~\bibnamefont {Neuhauser}},\ }\bibfield  {title} {\bibinfo
  {title} {Swift {GW} beyond 10,000 electrons using sparse stochastic
  compression},\ }\href {https://doi.org/10.1103/PhysRevB.98.075107} {\bibfield
   {journal} {\bibinfo  {journal} {Phys. Rev. B}\ }\textbf {\bibinfo {volume}
  {98}},\ \bibinfo {pages} {075107} (\bibinfo {year} {2018})}\BibitemShut
  {NoStop}%
\bibitem [{\citenamefont {Ben}\ \emph {et~al.}(2020)\citenamefont {Ben},
  \citenamefont {Yang}, \citenamefont {Li}, \citenamefont {Jornada},
  \citenamefont {Louie},\ and\ \citenamefont {Deslippe}}]{Mauro2020}%
  \BibitemOpen
  \bibfield  {author} {\bibinfo {author} {\bibfnamefont {M.~D.}\ \bibnamefont
  {Ben}}, \bibinfo {author} {\bibfnamefont {C.}~\bibnamefont {Yang}}, \bibinfo
  {author} {\bibfnamefont {Z.}~\bibnamefont {Li}}, \bibinfo {author}
  {\bibfnamefont {F.~H.~d.}\ \bibnamefont {Jornada}}, \bibinfo {author}
  {\bibfnamefont {S.~G.}\ \bibnamefont {Louie}},\ and\ \bibinfo {author}
  {\bibfnamefont {J.}~\bibnamefont {Deslippe}},\ }\bibfield  {title} {\bibinfo
  {title} {Accelerating large-scale excited-state gw calculations on leadership
  {HPC} systems},\ }in\ \href {https://doi.org/10.1109/SC41405.2020.00008}
  {\emph {\bibinfo {booktitle} {SC20: International Conference for High
  Performance Computing, Networking, Storage and Analysis}}}\ (\bibinfo {year}
  {2020})\ pp.\ \bibinfo {pages} {1--11}\BibitemShut {NoStop}%
\bibitem [{\citenamefont {Duchemin}\ and\ \citenamefont
  {Blase}(2021)}]{duchemin2021_cubicscale}%
  \BibitemOpen
  \bibfield  {author} {\bibinfo {author} {\bibfnamefont {I.}~\bibnamefont
  {Duchemin}}\ and\ \bibinfo {author} {\bibfnamefont {X.}~\bibnamefont
  {Blase}},\ }\bibfield  {title} {\bibinfo {title} {Cubic-scaling all-electron
  {GW} calculations with a separable density-fitting space–time approach},\
  }\href {https://doi.org/10.1021/acs.jctc.1c00101} {\bibfield  {journal}
  {\bibinfo  {journal} {Journal of Chemical Theory and Computation}\ }\textbf
  {\bibinfo {volume} {17}},\ \bibinfo {pages} {2383} (\bibinfo {year}
  {2021})}\BibitemShut {NoStop}%
\bibitem [{\citenamefont {Yu}\ and\ \citenamefont {Govoni}(2022)}]{victor2022}%
  \BibitemOpen
  \bibfield  {author} {\bibinfo {author} {\bibfnamefont {V.~W.-z.}\
  \bibnamefont {Yu}}\ and\ \bibinfo {author} {\bibfnamefont {M.}~\bibnamefont
  {Govoni}},\ }\bibfield  {title} {\bibinfo {title} {{GPU} acceleration of
  large-scale full-frequency {GW} calculations},\ }\href
  {https://doi.org/10.1021/acs.jctc.2c00241} {\bibfield  {journal} {\bibinfo
  {journal} {Journal of Chemical Theory and Computation}\ }\textbf {\bibinfo
  {volume} {18}},\ \bibinfo {pages} {4690} (\bibinfo {year}
  {2022})}\BibitemShut {NoStop}%
\bibitem [{\citenamefont {Deslippe}\ \emph {et~al.}(2012)\citenamefont
  {Deslippe}, \citenamefont {Samsonidze}, \citenamefont {Strubbe},
  \citenamefont {Jain}, \citenamefont {Cohen},\ and\ \citenamefont
  {Louie}}]{DESLIPPE20121269}%
  \BibitemOpen
  \bibfield  {author} {\bibinfo {author} {\bibfnamefont {J.}~\bibnamefont
  {Deslippe}}, \bibinfo {author} {\bibfnamefont {G.}~\bibnamefont
  {Samsonidze}}, \bibinfo {author} {\bibfnamefont {D.~A.}\ \bibnamefont
  {Strubbe}}, \bibinfo {author} {\bibfnamefont {M.}~\bibnamefont {Jain}},
  \bibinfo {author} {\bibfnamefont {M.~L.}\ \bibnamefont {Cohen}},\ and\
  \bibinfo {author} {\bibfnamefont {S.~G.}\ \bibnamefont {Louie}},\ }\bibfield
  {title} {\bibinfo {title} {\texttt{BerkeleyGW}: A massively parallel computer
  package for the calculation of the quasiparticle and optical properties of
  materials and nanostructures},\ }\href
  {https://doi.org/https://doi.org/10.1016/j.cpc.2011.12.006} {\bibfield
  {journal} {\bibinfo  {journal} {Computer Physics Communications}\ }\textbf
  {\bibinfo {volume} {183}},\ \bibinfo {pages} {1269} (\bibinfo {year}
  {2012})}\BibitemShut {NoStop}%
\bibitem [{\citenamefont {Sangalli}\ \emph {et~al.}(2019)\citenamefont
  {Sangalli}, \citenamefont {Ferretti}, \citenamefont {Miranda}, \citenamefont
  {Attaccalite}, \citenamefont {Marri}, \citenamefont {Cannuccia},
  \citenamefont {Melo}, \citenamefont {Marsili}, \citenamefont {Paleari},
  \citenamefont {Marrazzo}, \citenamefont {Prandini}, \citenamefont
  {Bonf{\`{a}}}, \citenamefont {Atambo}, \citenamefont {Affinito},
  \citenamefont {Palummo}, \citenamefont {Molina-S{\'{a}}nchez}, \citenamefont
  {Hogan}, \citenamefont {Grüning}, \citenamefont {Varsano},\ and\
  \citenamefont {Marini}}]{Sangalli_2019}%
  \BibitemOpen
  \bibfield  {author} {\bibinfo {author} {\bibfnamefont {D.}~\bibnamefont
  {Sangalli}}, \bibinfo {author} {\bibfnamefont {A.}~\bibnamefont {Ferretti}},
  \bibinfo {author} {\bibfnamefont {H.}~\bibnamefont {Miranda}}, \bibinfo
  {author} {\bibfnamefont {C.}~\bibnamefont {Attaccalite}}, \bibinfo {author}
  {\bibfnamefont {I.}~\bibnamefont {Marri}}, \bibinfo {author} {\bibfnamefont
  {E.}~\bibnamefont {Cannuccia}}, \bibinfo {author} {\bibfnamefont
  {P.}~\bibnamefont {Melo}}, \bibinfo {author} {\bibfnamefont {M.}~\bibnamefont
  {Marsili}}, \bibinfo {author} {\bibfnamefont {F.}~\bibnamefont {Paleari}},
  \bibinfo {author} {\bibfnamefont {A.}~\bibnamefont {Marrazzo}}, \bibinfo
  {author} {\bibfnamefont {G.}~\bibnamefont {Prandini}}, \bibinfo {author}
  {\bibfnamefont {P.}~\bibnamefont {Bonf{\`{a}}}}, \bibinfo {author}
  {\bibfnamefont {M.~O.}\ \bibnamefont {Atambo}}, \bibinfo {author}
  {\bibfnamefont {F.}~\bibnamefont {Affinito}}, \bibinfo {author}
  {\bibfnamefont {M.}~\bibnamefont {Palummo}}, \bibinfo {author} {\bibfnamefont
  {A.}~\bibnamefont {Molina-S{\'{a}}nchez}}, \bibinfo {author} {\bibfnamefont
  {C.}~\bibnamefont {Hogan}}, \bibinfo {author} {\bibfnamefont
  {M.}~\bibnamefont {Grüning}}, \bibinfo {author} {\bibfnamefont
  {D.}~\bibnamefont {Varsano}},\ and\ \bibinfo {author} {\bibfnamefont
  {A.}~\bibnamefont {Marini}},\ }\bibfield  {title} {\bibinfo {title}
  {Many-body perturbation theory calculations using the yambo code},\ }\href
  {https://doi.org/10.1088/1361-648x/ab15d0} {\bibfield  {journal} {\bibinfo
  {journal} {Journal of Physics: Condensed Matter}\ }\textbf {\bibinfo {volume}
  {31}},\ \bibinfo {pages} {325902} (\bibinfo {year} {2019})}\BibitemShut
  {NoStop}%
\bibitem [{\citenamefont {Hedin}(1999)}]{Hedin_1999}%
  \BibitemOpen
  \bibfield  {author} {\bibinfo {author} {\bibfnamefont {L.}~\bibnamefont
  {Hedin}},\ }\bibfield  {title} {\bibinfo {title} {On correlation effects in
  electron spectroscopies and the {GW} approximation},\ }\href
  {https://doi.org/10.1088/0953-8984/11/42/201} {\bibfield  {journal} {\bibinfo
   {journal} {Journal of Physics: Condensed Matter}\ }\textbf {\bibinfo
  {volume} {11}},\ \bibinfo {pages} {R489} (\bibinfo {year}
  {1999})}\BibitemShut {NoStop}%
\bibitem [{\citenamefont {Casida}(1995)}]{casida1995}%
  \BibitemOpen
  \bibfield  {author} {\bibinfo {author} {\bibfnamefont {M.~E.}\ \bibnamefont
  {Casida}},\ }\bibfield  {title} {\bibinfo {title} {Time-dependent density
  functional response theory for molecules},\ }in\ \href
  {https://doi.org/10.1142/9789812830586_0005} {\emph {\bibinfo {booktitle}
  {Recent Advances in Density Functional Methods}}},\ \bibinfo {editor} {edited
  by\ \bibinfo {editor} {\bibfnamefont {D.~P.}\ \bibnamefont {Chong}}}\
  (\bibinfo  {publisher} {World Scientific},\ \bibinfo {year} {1995})\ pp.\
  \bibinfo {pages} {155--192}\BibitemShut {NoStop}%
\bibitem [{\citenamefont {Bruneval}\ \emph {et~al.}(2016)\citenamefont
  {Bruneval}, \citenamefont {Rangel}, \citenamefont {Hamed}, \citenamefont
  {Shao}, \citenamefont {Yang},\ and\ \citenamefont
  {Neaton}}]{BRUNEVAL2016149}%
  \BibitemOpen
  \bibfield  {author} {\bibinfo {author} {\bibfnamefont {F.}~\bibnamefont
  {Bruneval}}, \bibinfo {author} {\bibfnamefont {T.}~\bibnamefont {Rangel}},
  \bibinfo {author} {\bibfnamefont {S.~M.}\ \bibnamefont {Hamed}}, \bibinfo
  {author} {\bibfnamefont {M.}~\bibnamefont {Shao}}, \bibinfo {author}
  {\bibfnamefont {C.}~\bibnamefont {Yang}},\ and\ \bibinfo {author}
  {\bibfnamefont {J.~B.}\ \bibnamefont {Neaton}},\ }\bibfield  {title}
  {\bibinfo {title} {\texttt{MOLGW} 1: Many-body perturbation theory software
  for atoms, molecules, and clusters},\ }\href
  {https://doi.org/https://doi.org/10.1016/j.cpc.2016.06.019} {\bibfield
  {journal} {\bibinfo  {journal} {Computer Physics Communications}\ }\textbf
  {\bibinfo {volume} {208}},\ \bibinfo {pages} {149} (\bibinfo {year}
  {2016})}\BibitemShut {NoStop}%
\bibitem [{\citenamefont {Tiago}\ and\ \citenamefont
  {Chelikowsky}(2006)}]{Tiago2006}%
  \BibitemOpen
  \bibfield  {author} {\bibinfo {author} {\bibfnamefont {M.~L.}\ \bibnamefont
  {Tiago}}\ and\ \bibinfo {author} {\bibfnamefont {J.~R.}\ \bibnamefont
  {Chelikowsky}},\ }\bibfield  {title} {\bibinfo {title} {Optical excitations
  in organic molecules, clusters, and defects studied by first-principles
  green's function methods},\ }\href
  {https://doi.org/10.1103/PhysRevB.73.205334} {\bibfield  {journal} {\bibinfo
  {journal} {Phys. Rev. B}\ }\textbf {\bibinfo {volume} {73}},\ \bibinfo
  {pages} {205334} (\bibinfo {year} {2006})}\BibitemShut {NoStop}%
\bibitem [{\citenamefont {van Setten}\ \emph {et~al.}(2013)\citenamefont {van
  Setten}, \citenamefont {Weigend},\ and\ \citenamefont
  {Evers}}]{vansetten2013}%
  \BibitemOpen
  \bibfield  {author} {\bibinfo {author} {\bibfnamefont {M.~J.}\ \bibnamefont
  {van Setten}}, \bibinfo {author} {\bibfnamefont {F.}~\bibnamefont
  {Weigend}},\ and\ \bibinfo {author} {\bibfnamefont {F.}~\bibnamefont
  {Evers}},\ }\bibfield  {title} {\bibinfo {title} {The gw-method for quantum
  chemistry applications: Theory and implementation},\ }\href
  {https://doi.org/10.1021/ct300648t} {\bibfield  {journal} {\bibinfo
  {journal} {Journal of Chemical Theory and Computation}\ }\textbf {\bibinfo
  {volume} {9}},\ \bibinfo {pages} {232} (\bibinfo {year} {2013})}\BibitemShut
  {NoStop}%
\bibitem [{\citenamefont {Mejia-Rodriguez}\ \emph {et~al.}(2021)\citenamefont
  {Mejia-Rodriguez}, \citenamefont {Kunitsa}, \citenamefont {Aprà},\ and\
  \citenamefont {Govind}}]{daniel2021}%
  \BibitemOpen
  \bibfield  {author} {\bibinfo {author} {\bibfnamefont {D.}~\bibnamefont
  {Mejia-Rodriguez}}, \bibinfo {author} {\bibfnamefont {A.}~\bibnamefont
  {Kunitsa}}, \bibinfo {author} {\bibfnamefont {E.}~\bibnamefont {Aprà}},\
  and\ \bibinfo {author} {\bibfnamefont {N.}~\bibnamefont {Govind}},\
  }\bibfield  {title} {\bibinfo {title} {Scalable molecular {GW} calculations:
  Valence and core spectra},\ }\href {https://doi.org/10.1021/acs.jctc.1c00738}
  {\bibfield  {journal} {\bibinfo  {journal} {Journal of Chemical Theory and
  Computation}\ }\textbf {\bibinfo {volume} {17}},\ \bibinfo {pages} {7504}
  (\bibinfo {year} {2021})}\BibitemShut {NoStop}%
\bibitem [{\citenamefont {Onida}\ \emph {et~al.}(2002)\citenamefont {Onida},
  \citenamefont {Reining},\ and\ \citenamefont {Rubio}}]{Onida2002}%
  \BibitemOpen
  \bibfield  {author} {\bibinfo {author} {\bibfnamefont {G.}~\bibnamefont
  {Onida}}, \bibinfo {author} {\bibfnamefont {L.}~\bibnamefont {Reining}},\
  and\ \bibinfo {author} {\bibfnamefont {A.}~\bibnamefont {Rubio}},\ }\bibfield
   {title} {\bibinfo {title} {Electronic excitations: {{Density}-functional}
  versus many-body green’s-function approaches},\ }\href
  {https://doi.org/10.1103/revmodphys.74.601} {\bibfield  {journal} {\bibinfo
  {journal} {Rev. Mod. Phys.}\ }\textbf {\bibinfo {volume} {74}},\ \bibinfo
  {pages} {601} (\bibinfo {year} {2002})}\BibitemShut {NoStop}%
\bibitem [{sup()}]{supp}%
  \BibitemOpen
  \href@noop {} {}\bibinfo {note} {See Supplemental Material at
  [URL-will-be-inserted-by-publisher] for the details of Eq. (9) - (12),
  computational details, HOMO-LUMO gaps of silicon nanoclusters, and analysis
  of the scaling of computation costs.}\BibitemShut {Stop}%
\bibitem [{\citenamefont {Haydock}(1980)}]{HAYDOCK198011}%
  \BibitemOpen
  \bibfield  {author} {\bibinfo {author} {\bibfnamefont {R.}~\bibnamefont
  {Haydock}},\ }\bibfield  {title} {\bibinfo {title} {The recursive solution of
  the schrödinger equation},\ }\href
  {https://doi.org/https://doi.org/10.1016/0010-4655(80)90101-0} {\bibfield
  {journal} {\bibinfo  {journal} {Computer Physics Communications}\ }\textbf
  {\bibinfo {volume} {20}},\ \bibinfo {pages} {11} (\bibinfo {year}
  {1980})}\BibitemShut {NoStop}%
\bibitem [{\citenamefont {Meyer}\ and\ \citenamefont {Pal}(1989)}]{meyer1989}%
  \BibitemOpen
  \bibfield  {author} {\bibinfo {author} {\bibfnamefont {H.}~\bibnamefont
  {Meyer}}\ and\ \bibinfo {author} {\bibfnamefont {S.}~\bibnamefont {Pal}},\
  }\bibfield  {title} {\bibinfo {title} {{A band‐Lanczos method for computing
  matrix elements of a resolvent}},\ }\href {https://doi.org/10.1063/1.457438}
  {\bibfield  {journal} {\bibinfo  {journal} {The Journal of Chemical Physics}\
  }\textbf {\bibinfo {volume} {91}},\ \bibinfo {pages} {6195} (\bibinfo {year}
  {1989})}\BibitemShut {NoStop}%
\bibitem [{\citenamefont {Malcıoğlu}\ \emph {et~al.}(2011)\citenamefont
  {Malcıoğlu}, \citenamefont {Gebauer}, \citenamefont {Rocca},\ and\
  \citenamefont {Baroni}}]{MALCIOGLU20111744}%
  \BibitemOpen
  \bibfield  {author} {\bibinfo {author} {\bibfnamefont {O.~B.}\ \bibnamefont
  {Malcıoğlu}}, \bibinfo {author} {\bibfnamefont {R.}~\bibnamefont
  {Gebauer}}, \bibinfo {author} {\bibfnamefont {D.}~\bibnamefont {Rocca}},\
  and\ \bibinfo {author} {\bibfnamefont {S.}~\bibnamefont {Baroni}},\
  }\bibfield  {title} {\bibinfo {title} {turbotddft – a code for the
  simulation of molecular spectra using the liouville–lanczos approach to
  time-dependent density-functional perturbation theory},\ }\href
  {https://doi.org/https://doi.org/10.1016/j.cpc.2011.04.020} {\bibfield
  {journal} {\bibinfo  {journal} {Computer Physics Communications}\ }\textbf
  {\bibinfo {volume} {182}},\ \bibinfo {pages} {1744} (\bibinfo {year}
  {2011})}\BibitemShut {NoStop}%
\bibitem [{\citenamefont {Walker}\ \emph {et~al.}(2006)\citenamefont {Walker},
  \citenamefont {Saitta}, \citenamefont {Gebauer},\ and\ \citenamefont
  {Baroni}}]{walker2006}%
  \BibitemOpen
  \bibfield  {author} {\bibinfo {author} {\bibfnamefont {B.}~\bibnamefont
  {Walker}}, \bibinfo {author} {\bibfnamefont {A.~M.}\ \bibnamefont {Saitta}},
  \bibinfo {author} {\bibfnamefont {R.}~\bibnamefont {Gebauer}},\ and\ \bibinfo
  {author} {\bibfnamefont {S.}~\bibnamefont {Baroni}},\ }\bibfield  {title}
  {\bibinfo {title} {Efficient approach to time-dependent density-functional
  perturbation theory for optical spectroscopy},\ }\href
  {https://doi.org/10.1103/PhysRevLett.96.113001} {\bibfield  {journal}
  {\bibinfo  {journal} {Phys. Rev. Lett.}\ }\textbf {\bibinfo {volume} {96}},\
  \bibinfo {pages} {113001} (\bibinfo {year} {2006})}\BibitemShut {NoStop}%
\bibitem [{\citenamefont {Zamok}\ \emph {et~al.}(2022)\citenamefont {Zamok},
  \citenamefont {Coriani},\ and\ \citenamefont {Sauer}}]{zamok2022}%
  \BibitemOpen
  \bibfield  {author} {\bibinfo {author} {\bibfnamefont {L.}~\bibnamefont
  {Zamok}}, \bibinfo {author} {\bibfnamefont {S.}~\bibnamefont {Coriani}},\
  and\ \bibinfo {author} {\bibfnamefont {S.~P.~A.}\ \bibnamefont {Sauer}},\
  }\bibfield  {title} {\bibinfo {title} {{A tale of two vectors: A Lanczos
  algorithm for calculating RPA mean excitation energies}},\ }\href
  {https://doi.org/10.1063/5.0071144} {\bibfield  {journal} {\bibinfo
  {journal} {The Journal of Chemical Physics}\ }\textbf {\bibinfo {volume}
  {156}},\ \bibinfo {pages} {014102} (\bibinfo {year} {2022})}\BibitemShut
  {NoStop}%
\bibitem [{\citenamefont {Grüning}\ \emph {et~al.}(2011)\citenamefont
  {Grüning}, \citenamefont {Marini},\ and\ \citenamefont
  {Gonze}}]{GRUNING20112148}%
  \BibitemOpen
  \bibfield  {author} {\bibinfo {author} {\bibfnamefont {M.}~\bibnamefont
  {Grüning}}, \bibinfo {author} {\bibfnamefont {A.}~\bibnamefont {Marini}},\
  and\ \bibinfo {author} {\bibfnamefont {X.}~\bibnamefont {Gonze}},\ }\bibfield
   {title} {\bibinfo {title} {Implementation and testing of lanczos-based
  algorithms for random-phase approximation eigenproblems},\ }\href
  {https://doi.org/https://doi.org/10.1016/j.commatsci.2011.02.021} {\bibfield
  {journal} {\bibinfo  {journal} {Computational Materials Science}\ }\textbf
  {\bibinfo {volume} {50}},\ \bibinfo {pages} {2148} (\bibinfo {year}
  {2011})}\BibitemShut {NoStop}%
\bibitem [{\citenamefont {Benedict}\ \emph {et~al.}(1998)\citenamefont
  {Benedict}, \citenamefont {Shirley},\ and\ \citenamefont
  {Bohn}}]{Benedict1998}%
  \BibitemOpen
  \bibfield  {author} {\bibinfo {author} {\bibfnamefont {L.~X.}\ \bibnamefont
  {Benedict}}, \bibinfo {author} {\bibfnamefont {E.~L.}\ \bibnamefont
  {Shirley}},\ and\ \bibinfo {author} {\bibfnamefont {R.~B.}\ \bibnamefont
  {Bohn}},\ }\bibfield  {title} {\bibinfo {title} {Theory of optical absorption
  in diamond, {Si}, {Ge}, and {GaAs}},\ }\href
  {https://doi.org/10.1103/PhysRevB.57.R9385} {\bibfield  {journal} {\bibinfo
  {journal} {Phys. Rev. B}\ }\textbf {\bibinfo {volume} {57}},\ \bibinfo
  {pages} {R9385} (\bibinfo {year} {1998})}\BibitemShut {NoStop}%
\bibitem [{\citenamefont {Shao}\ \emph {et~al.}(2018)\citenamefont {Shao},
  \citenamefont {da~Jornada}, \citenamefont {Lin}, \citenamefont {Yang},
  \citenamefont {Deslippe},\ and\ \citenamefont {Louie}}]{shao2018}%
  \BibitemOpen
  \bibfield  {author} {\bibinfo {author} {\bibfnamefont {M.}~\bibnamefont
  {Shao}}, \bibinfo {author} {\bibfnamefont {F.~H.}\ \bibnamefont
  {da~Jornada}}, \bibinfo {author} {\bibfnamefont {L.}~\bibnamefont {Lin}},
  \bibinfo {author} {\bibfnamefont {C.}~\bibnamefont {Yang}}, \bibinfo {author}
  {\bibfnamefont {J.}~\bibnamefont {Deslippe}},\ and\ \bibinfo {author}
  {\bibfnamefont {S.~G.}\ \bibnamefont {Louie}},\ }\bibfield  {title} {\bibinfo
  {title} {A structure preserving lanczos algorithm for computing the optical
  absorption spectrum},\ }\href {https://doi.org/10.1137/16M1102641} {\bibfield
   {journal} {\bibinfo  {journal} {SIAM Journal on Matrix Analysis and
  Applications}\ }\textbf {\bibinfo {volume} {39}},\ \bibinfo {pages} {683}
  (\bibinfo {year} {2018})}\BibitemShut {NoStop}%
\bibitem [{\citenamefont {Govoni}\ and\ \citenamefont
  {Galli}(2015)}]{govoni2015}%
  \BibitemOpen
  \bibfield  {author} {\bibinfo {author} {\bibfnamefont {M.}~\bibnamefont
  {Govoni}}\ and\ \bibinfo {author} {\bibfnamefont {G.}~\bibnamefont {Galli}},\
  }\bibfield  {title} {\bibinfo {title} {Large scale {GW} calculations},\
  }\href {https://doi.org/10.1021/ct500958p} {\bibfield  {journal} {\bibinfo
  {journal} {Journal of Chemical Theory and Computation}\ }\textbf {\bibinfo
  {volume} {11}},\ \bibinfo {pages} {2680} (\bibinfo {year}
  {2015})}\BibitemShut {NoStop}%
\bibitem [{\citenamefont {Umari}\ \emph {et~al.}(2010)\citenamefont {Umari},
  \citenamefont {Stenuit},\ and\ \citenamefont {Baroni}}]{Umari2010}%
  \BibitemOpen
  \bibfield  {author} {\bibinfo {author} {\bibfnamefont {P.}~\bibnamefont
  {Umari}}, \bibinfo {author} {\bibfnamefont {G.}~\bibnamefont {Stenuit}},\
  and\ \bibinfo {author} {\bibfnamefont {S.}~\bibnamefont {Baroni}},\
  }\bibfield  {title} {\bibinfo {title} {{GW} quasiparticle spectra from
  occupied states only},\ }\href {https://doi.org/10.1103/PhysRevB.81.115104}
  {\bibfield  {journal} {\bibinfo  {journal} {Phys. Rev. B}\ }\textbf {\bibinfo
  {volume} {81}},\ \bibinfo {pages} {115104} (\bibinfo {year}
  {2010})}\BibitemShut {NoStop}%
\bibitem [{\citenamefont {Laflamme~Janssen}\ \emph {et~al.}(2015)\citenamefont
  {Laflamme~Janssen}, \citenamefont {Rousseau},\ and\ \citenamefont
  {C\^ot\'e}}]{laflamme2015}%
  \BibitemOpen
  \bibfield  {author} {\bibinfo {author} {\bibfnamefont {J.}~\bibnamefont
  {Laflamme~Janssen}}, \bibinfo {author} {\bibfnamefont {B.}~\bibnamefont
  {Rousseau}},\ and\ \bibinfo {author} {\bibfnamefont {M.}~\bibnamefont
  {C\^ot\'e}},\ }\bibfield  {title} {\bibinfo {title} {Efficient dielectric
  matrix calculations using the lanczos algorithm for fast many-body
  ${G}_{0}{W}_{0}$ implementations},\ }\href
  {https://doi.org/10.1103/PhysRevB.91.125120} {\bibfield  {journal} {\bibinfo
  {journal} {Phys. Rev. B}\ }\textbf {\bibinfo {volume} {91}},\ \bibinfo
  {pages} {125120} (\bibinfo {year} {2015})}\BibitemShut {NoStop}%
\bibitem [{\citenamefont {Bintrim}\ and\ \citenamefont
  {Berkelbach}(2021)}]{bintrim2021}%
  \BibitemOpen
  \bibfield  {author} {\bibinfo {author} {\bibfnamefont {S.~J.}\ \bibnamefont
  {Bintrim}}\ and\ \bibinfo {author} {\bibfnamefont {T.~C.}\ \bibnamefont
  {Berkelbach}},\ }\bibfield  {title} {\bibinfo {title} {{Full-frequency GW
  without frequency}},\ }\href {https://doi.org/10.1063/5.0035141} {\bibfield
  {journal} {\bibinfo  {journal} {The Journal of Chemical Physics}\ }\textbf
  {\bibinfo {volume} {154}},\ \bibinfo {pages} {041101} (\bibinfo {year}
  {2021})}\BibitemShut {NoStop}%
\bibitem [{\citenamefont {Scott}\ \emph {et~al.}(2023)\citenamefont {Scott},
  \citenamefont {Backhouse},\ and\ \citenamefont {Booth}}]{scott2023}%
  \BibitemOpen
  \bibfield  {author} {\bibinfo {author} {\bibfnamefont {C.~J.~C.}\
  \bibnamefont {Scott}}, \bibinfo {author} {\bibfnamefont {O.~J.}\ \bibnamefont
  {Backhouse}},\ and\ \bibinfo {author} {\bibfnamefont {G.~H.}\ \bibnamefont
  {Booth}},\ }\bibfield  {title} {\bibinfo {title} {{A “moment-conserving”
  reformulation of GW theory}},\ }\href {https://doi.org/10.1063/5.0143291}
  {\bibfield  {journal} {\bibinfo  {journal} {The Journal of Chemical Physics}\
  }\textbf {\bibinfo {volume} {158}},\ \bibinfo {pages} {124102} (\bibinfo
  {year} {2023})}\BibitemShut {NoStop}%
\bibitem [{\citenamefont {Backhouse}\ \emph {et~al.}(2021)\citenamefont
  {Backhouse}, \citenamefont {Santana-Bonilla},\ and\ \citenamefont
  {Booth}}]{backhouse2021}%
  \BibitemOpen
  \bibfield  {author} {\bibinfo {author} {\bibfnamefont {O.~J.}\ \bibnamefont
  {Backhouse}}, \bibinfo {author} {\bibfnamefont {A.}~\bibnamefont
  {Santana-Bonilla}},\ and\ \bibinfo {author} {\bibfnamefont {G.~H.}\
  \bibnamefont {Booth}},\ }\bibfield  {title} {\bibinfo {title} {Scalable and
  predictive spectra of correlated molecules with moment truncated iterated
  perturbation theory},\ }\href {https://doi.org/10.1021/acs.jpclett.1c02383}
  {\bibfield  {journal} {\bibinfo  {journal} {The Journal of Physical Chemistry
  Letters}\ }\textbf {\bibinfo {volume} {12}},\ \bibinfo {pages} {7650}
  (\bibinfo {year} {2021})}\BibitemShut {NoStop}%
\bibitem [{\citenamefont {Chiarotti}\ \emph {et~al.}(2022)\citenamefont
  {Chiarotti}, \citenamefont {Marzari},\ and\ \citenamefont
  {Ferretti}}]{chiarotti2022}%
  \BibitemOpen
  \bibfield  {author} {\bibinfo {author} {\bibfnamefont {T.}~\bibnamefont
  {Chiarotti}}, \bibinfo {author} {\bibfnamefont {N.}~\bibnamefont {Marzari}},\
  and\ \bibinfo {author} {\bibfnamefont {A.}~\bibnamefont {Ferretti}},\
  }\bibfield  {title} {\bibinfo {title} {Unified green's function approach for
  spectral and thermodynamic properties from algorithmic inversion of dynamical
  potentials},\ }\href {https://doi.org/10.1103/PhysRevResearch.4.013242}
  {\bibfield  {journal} {\bibinfo  {journal} {Phys. Rev. Res.}\ }\textbf
  {\bibinfo {volume} {4}},\ \bibinfo {pages} {013242} (\bibinfo {year}
  {2022})}\BibitemShut {NoStop}%
\bibitem [{\citenamefont {Leon}\ \emph {et~al.}(2023)\citenamefont {Leon},
  \citenamefont {Ferretti}, \citenamefont {Varsano}, \citenamefont {Molinari},\
  and\ \citenamefont {Cardoso}}]{leon2023}%
  \BibitemOpen
  \bibfield  {author} {\bibinfo {author} {\bibfnamefont {D.~A.}\ \bibnamefont
  {Leon}}, \bibinfo {author} {\bibfnamefont {A.}~\bibnamefont {Ferretti}},
  \bibinfo {author} {\bibfnamefont {D.}~\bibnamefont {Varsano}}, \bibinfo
  {author} {\bibfnamefont {E.}~\bibnamefont {Molinari}},\ and\ \bibinfo
  {author} {\bibfnamefont {C.}~\bibnamefont {Cardoso}},\ }\bibfield  {title}
  {\bibinfo {title} {Efficient full frequency {GW} for metals using a multipole
  approach for the dielectric screening},\ }\href
  {https://doi.org/10.1103/PhysRevB.107.155130} {\bibfield  {journal} {\bibinfo
   {journal} {Phys. Rev. B}\ }\textbf {\bibinfo {volume} {107}},\ \bibinfo
  {pages} {155130} (\bibinfo {year} {2023})}\BibitemShut {NoStop}%
\bibitem [{\citenamefont {Bintrim}\ and\ \citenamefont
  {Berkelbach}(2022)}]{bintrim2022}%
  \BibitemOpen
  \bibfield  {author} {\bibinfo {author} {\bibfnamefont {S.~J.}\ \bibnamefont
  {Bintrim}}\ and\ \bibinfo {author} {\bibfnamefont {T.~C.}\ \bibnamefont
  {Berkelbach}},\ }\bibfield  {title} {\bibinfo {title} {{Full-frequency
  dynamical Bethe–Salpeter equation without frequency and a study of double
  excitations}},\ }\href {https://doi.org/10.1063/5.0074434} {\bibfield
  {journal} {\bibinfo  {journal} {The Journal of Chemical Physics}\ }\textbf
  {\bibinfo {volume} {156}},\ \bibinfo {pages} {044114} (\bibinfo {year}
  {2022})}\BibitemShut {NoStop}%
\bibitem [{\citenamefont {Schirmer}\ \emph {et~al.}(1983)\citenamefont
  {Schirmer}, \citenamefont {Cederbaum},\ and\ \citenamefont
  {Walter}}]{schirmer1983}%
  \BibitemOpen
  \bibfield  {author} {\bibinfo {author} {\bibfnamefont {J.}~\bibnamefont
  {Schirmer}}, \bibinfo {author} {\bibfnamefont {L.~S.}\ \bibnamefont
  {Cederbaum}},\ and\ \bibinfo {author} {\bibfnamefont {O.}~\bibnamefont
  {Walter}},\ }\bibfield  {title} {\bibinfo {title} {New approach to the
  one-particle green's function for finite fermi systems},\ }\href
  {https://doi.org/10.1103/PhysRevA.28.1237} {\bibfield  {journal} {\bibinfo
  {journal} {Phys. Rev. A}\ }\textbf {\bibinfo {volume} {28}},\ \bibinfo
  {pages} {1237} (\bibinfo {year} {1983})}\BibitemShut {NoStop}%
\bibitem [{\citenamefont {van Setten}\ \emph {et~al.}(2015)\citenamefont {van
  Setten}, \citenamefont {Caruso}, \citenamefont {Sharifzadeh}, \citenamefont
  {Ren}, \citenamefont {Scheffler}, \citenamefont {Liu}, \citenamefont
  {Lischner}, \citenamefont {Lin}, \citenamefont {Deslippe}, \citenamefont
  {Louie}, \citenamefont {Yang}, \citenamefont {Weigend}, \citenamefont
  {Neaton}, \citenamefont {Evers},\ and\ \citenamefont
  {Rinke}}]{vansetten2015}%
  \BibitemOpen
  \bibfield  {author} {\bibinfo {author} {\bibfnamefont {M.~J.}\ \bibnamefont
  {van Setten}}, \bibinfo {author} {\bibfnamefont {F.}~\bibnamefont {Caruso}},
  \bibinfo {author} {\bibfnamefont {S.}~\bibnamefont {Sharifzadeh}}, \bibinfo
  {author} {\bibfnamefont {X.}~\bibnamefont {Ren}}, \bibinfo {author}
  {\bibfnamefont {M.}~\bibnamefont {Scheffler}}, \bibinfo {author}
  {\bibfnamefont {F.}~\bibnamefont {Liu}}, \bibinfo {author} {\bibfnamefont
  {J.}~\bibnamefont {Lischner}}, \bibinfo {author} {\bibfnamefont
  {L.}~\bibnamefont {Lin}}, \bibinfo {author} {\bibfnamefont {J.~R.}\
  \bibnamefont {Deslippe}}, \bibinfo {author} {\bibfnamefont {S.~G.}\
  \bibnamefont {Louie}}, \bibinfo {author} {\bibfnamefont {C.}~\bibnamefont
  {Yang}}, \bibinfo {author} {\bibfnamefont {F.}~\bibnamefont {Weigend}},
  \bibinfo {author} {\bibfnamefont {J.~B.}\ \bibnamefont {Neaton}}, \bibinfo
  {author} {\bibfnamefont {F.}~\bibnamefont {Evers}},\ and\ \bibinfo {author}
  {\bibfnamefont {P.}~\bibnamefont {Rinke}},\ }\bibfield  {title} {\bibinfo
  {title} {{GW100}: Benchmarking {G}$_0${W}$_0$ for molecular systems},\ }\href
  {https://doi.org/10.1021/acs.jctc.5b00453} {\bibfield  {journal} {\bibinfo
  {journal} {Journal of Chemical Theory and Computation}\ }\textbf {\bibinfo
  {volume} {11}},\ \bibinfo {pages} {5665} (\bibinfo {year}
  {2015})}\BibitemShut {NoStop}%
\bibitem [{\citenamefont {Govoni}\ and\ \citenamefont
  {Galli}(2018)}]{govoni2018}%
  \BibitemOpen
  \bibfield  {author} {\bibinfo {author} {\bibfnamefont {M.}~\bibnamefont
  {Govoni}}\ and\ \bibinfo {author} {\bibfnamefont {G.}~\bibnamefont {Galli}},\
  }\bibfield  {title} {\bibinfo {title} {{GW100}: Comparison of methods and
  accuracy of results obtained with the west code},\ }\href
  {https://doi.org/10.1021/acs.jctc.7b00952} {\bibfield  {journal} {\bibinfo
  {journal} {Journal of Chemical Theory and Computation}\ }\textbf {\bibinfo
  {volume} {14}},\ \bibinfo {pages} {1895} (\bibinfo {year}
  {2018})}\BibitemShut {NoStop}%
\bibitem [{\citenamefont {Förster}\ and\ \citenamefont
  {Visscher}(2021)}]{forster2021}%
  \BibitemOpen
  \bibfield  {author} {\bibinfo {author} {\bibfnamefont {A.}~\bibnamefont
  {Förster}}\ and\ \bibinfo {author} {\bibfnamefont {L.}~\bibnamefont
  {Visscher}},\ }\bibfield  {title} {\bibinfo {title} {{GW100}: A slater-type
  orbital perspective},\ }\href {https://doi.org/10.1021/acs.jctc.1c00308}
  {\bibfield  {journal} {\bibinfo  {journal} {Journal of Chemical Theory and
  Computation}\ }\textbf {\bibinfo {volume} {17}},\ \bibinfo {pages} {5080}
  (\bibinfo {year} {2021})}\BibitemShut {NoStop}%
\bibitem [{\citenamefont {Gao}\ and\ \citenamefont
  {Chelikowsky}(2019)}]{gao2019_gw100}%
  \BibitemOpen
  \bibfield  {author} {\bibinfo {author} {\bibfnamefont {W.}~\bibnamefont
  {Gao}}\ and\ \bibinfo {author} {\bibfnamefont {J.~R.}\ \bibnamefont
  {Chelikowsky}},\ }\bibfield  {title} {\bibinfo {title} {Real-space based
  benchmark of {G}$_0${W}$_0$ calculations on {GW100}: Effects of semicore
  orbitals and orbital reordering},\ }\href
  {https://doi.org/10.1021/acs.jctc.9b00520} {\bibfield  {journal} {\bibinfo
  {journal} {Journal of Chemical Theory and Computation}\ }\textbf {\bibinfo
  {volume} {15}},\ \bibinfo {pages} {5299} (\bibinfo {year}
  {2019})}\BibitemShut {NoStop}%
\bibitem [{\citenamefont {Caruso}\ \emph {et~al.}(2013)\citenamefont {Caruso},
  \citenamefont {Rinke}, \citenamefont {Ren}, \citenamefont {Rubio},\ and\
  \citenamefont {Scheffler}}]{caruso2013}%
  \BibitemOpen
  \bibfield  {author} {\bibinfo {author} {\bibfnamefont {F.}~\bibnamefont
  {Caruso}}, \bibinfo {author} {\bibfnamefont {P.}~\bibnamefont {Rinke}},
  \bibinfo {author} {\bibfnamefont {X.}~\bibnamefont {Ren}}, \bibinfo {author}
  {\bibfnamefont {A.}~\bibnamefont {Rubio}},\ and\ \bibinfo {author}
  {\bibfnamefont {M.}~\bibnamefont {Scheffler}},\ }\bibfield  {title} {\bibinfo
  {title} {Self-consistent {GW}: All-electron implementation with localized
  basis functions},\ }\href {https://doi.org/10.1103/PhysRevB.88.075105}
  {\bibfield  {journal} {\bibinfo  {journal} {Phys. Rev. B}\ }\textbf {\bibinfo
  {volume} {88}},\ \bibinfo {pages} {075105} (\bibinfo {year}
  {2013})}\BibitemShut {NoStop}%
\bibitem [{\citenamefont {Marom}\ \emph {et~al.}(2012)\citenamefont {Marom},
  \citenamefont {Caruso}, \citenamefont {Ren}, \citenamefont {Hofmann},
  \citenamefont {K\"orzd\"orfer}, \citenamefont {Chelikowsky}, \citenamefont
  {Rubio}, \citenamefont {Scheffler},\ and\ \citenamefont {Rinke}}]{marom2012}%
  \BibitemOpen
  \bibfield  {author} {\bibinfo {author} {\bibfnamefont {N.}~\bibnamefont
  {Marom}}, \bibinfo {author} {\bibfnamefont {F.}~\bibnamefont {Caruso}},
  \bibinfo {author} {\bibfnamefont {X.}~\bibnamefont {Ren}}, \bibinfo {author}
  {\bibfnamefont {O.~T.}\ \bibnamefont {Hofmann}}, \bibinfo {author}
  {\bibfnamefont {T.}~\bibnamefont {K\"orzd\"orfer}}, \bibinfo {author}
  {\bibfnamefont {J.~R.}\ \bibnamefont {Chelikowsky}}, \bibinfo {author}
  {\bibfnamefont {A.}~\bibnamefont {Rubio}}, \bibinfo {author} {\bibfnamefont
  {M.}~\bibnamefont {Scheffler}},\ and\ \bibinfo {author} {\bibfnamefont
  {P.}~\bibnamefont {Rinke}},\ }\bibfield  {title} {\bibinfo {title} {Benchmark
  of {GW} methods for azabenzenes},\ }\href
  {https://doi.org/10.1103/PhysRevB.86.245127} {\bibfield  {journal} {\bibinfo
  {journal} {Phys. Rev. B}\ }\textbf {\bibinfo {volume} {86}},\ \bibinfo
  {pages} {245127} (\bibinfo {year} {2012})}\BibitemShut {NoStop}%
\bibitem [{\citenamefont {Rinke}\ \emph {et~al.}(2005)\citenamefont {Rinke},
  \citenamefont {Qteish}, \citenamefont {Neugebauer}, \citenamefont
  {Freysoldt},\ and\ \citenamefont {Scheffler}}]{Rinke_2005}%
  \BibitemOpen
  \bibfield  {author} {\bibinfo {author} {\bibfnamefont {P.}~\bibnamefont
  {Rinke}}, \bibinfo {author} {\bibfnamefont {A.}~\bibnamefont {Qteish}},
  \bibinfo {author} {\bibfnamefont {J.}~\bibnamefont {Neugebauer}}, \bibinfo
  {author} {\bibfnamefont {C.}~\bibnamefont {Freysoldt}},\ and\ \bibinfo
  {author} {\bibfnamefont {M.}~\bibnamefont {Scheffler}},\ }\bibfield  {title}
  {\bibinfo {title} {Combining {GW} calculations with exact-exchange
  density-functional theory: an analysis of valence-band photoemission for
  compound semiconductors},\ }\href {https://doi.org/10.1088/1367-2630/7/1/126}
  {\bibfield  {journal} {\bibinfo  {journal} {New Journal of Physics}\ }\textbf
  {\bibinfo {volume} {7}},\ \bibinfo {pages} {126} (\bibinfo {year}
  {2005})}\BibitemShut {NoStop}%
\bibitem [{\citenamefont {Dauth}\ \emph {et~al.}(2016)\citenamefont {Dauth},
  \citenamefont {Caruso}, \citenamefont {K\"ummel},\ and\ \citenamefont
  {Rinke}}]{Dauth2016}%
  \BibitemOpen
  \bibfield  {author} {\bibinfo {author} {\bibfnamefont {M.}~\bibnamefont
  {Dauth}}, \bibinfo {author} {\bibfnamefont {F.}~\bibnamefont {Caruso}},
  \bibinfo {author} {\bibfnamefont {S.}~\bibnamefont {K\"ummel}},\ and\
  \bibinfo {author} {\bibfnamefont {P.}~\bibnamefont {Rinke}},\ }\bibfield
  {title} {\bibinfo {title} {Piecewise linearity in the ${GW}$ approximation
  for accurate quasiparticle energy predictions},\ }\href
  {https://doi.org/10.1103/PhysRevB.93.121115} {\bibfield  {journal} {\bibinfo
  {journal} {Phys. Rev. B}\ }\textbf {\bibinfo {volume} {93}},\ \bibinfo
  {pages} {121115} (\bibinfo {year} {2016})}\BibitemShut {NoStop}%
\bibitem [{\citenamefont {Hung}\ \emph {et~al.}(2016)\citenamefont {Hung},
  \citenamefont {da~Jornada}, \citenamefont {Souto-Casares}, \citenamefont
  {Chelikowsky}, \citenamefont {Louie},\ and\ \citenamefont
  {\"O\ifmmode~\breve{g}\else \u{g}\fi{}\"ut}}]{Hung2016}%
  \BibitemOpen
  \bibfield  {author} {\bibinfo {author} {\bibfnamefont {L.}~\bibnamefont
  {Hung}}, \bibinfo {author} {\bibfnamefont {F.~H.}\ \bibnamefont
  {da~Jornada}}, \bibinfo {author} {\bibfnamefont {J.}~\bibnamefont
  {Souto-Casares}}, \bibinfo {author} {\bibfnamefont {J.~R.}\ \bibnamefont
  {Chelikowsky}}, \bibinfo {author} {\bibfnamefont {S.~G.}\ \bibnamefont
  {Louie}},\ and\ \bibinfo {author} {\bibfnamefont {S.}~\bibnamefont
  {\"O\ifmmode~\breve{g}\else \u{g}\fi{}\"ut}},\ }\bibfield  {title} {\bibinfo
  {title} {Excitation spectra of aromatic molecules within a real-space
  ${GW}$-{BSE} formalism: Role of self-consistency and vertex corrections},\
  }\href {https://doi.org/10.1103/PhysRevB.94.085125} {\bibfield  {journal}
  {\bibinfo  {journal} {Phys. Rev. B}\ }\textbf {\bibinfo {volume} {94}},\
  \bibinfo {pages} {085125} (\bibinfo {year} {2016})}\BibitemShut {NoStop}%
\bibitem [{\citenamefont {Kronik}\ \emph {et~al.}(2006)\citenamefont {Kronik},
  \citenamefont {Makmal}, \citenamefont {Tiago}, \citenamefont {Alemany},
  \citenamefont {Jain}, \citenamefont {Huang}, \citenamefont {Saad},\ and\
  \citenamefont {Chelikowsky}}]{Kronik2006}%
  \BibitemOpen
  \bibfield  {author} {\bibinfo {author} {\bibfnamefont {L.}~\bibnamefont
  {Kronik}}, \bibinfo {author} {\bibfnamefont {A.}~\bibnamefont {Makmal}},
  \bibinfo {author} {\bibfnamefont {M.~L.}\ \bibnamefont {Tiago}}, \bibinfo
  {author} {\bibfnamefont {M.~M.~G.}\ \bibnamefont {Alemany}}, \bibinfo
  {author} {\bibfnamefont {M.}~\bibnamefont {Jain}}, \bibinfo {author}
  {\bibfnamefont {X.}~\bibnamefont {Huang}}, \bibinfo {author} {\bibfnamefont
  {Y.}~\bibnamefont {Saad}},\ and\ \bibinfo {author} {\bibfnamefont {J.~R.}\
  \bibnamefont {Chelikowsky}},\ }\bibfield  {title} {\bibinfo {title} {{PARSEC
  – the pseudopotential algorithm for real-space electronic structure
  calculations: recent advances and novel applications to nano-structures}},\
  }\href {https://doi.org/10.1002/pssb.200541463} {\bibfield  {journal}
  {\bibinfo  {journal} {Phys. Status Solidi B}\ }\textbf {\bibinfo {volume}
  {243}},\ \bibinfo {pages} {1063} (\bibinfo {year} {2006})}\BibitemShut
  {NoStop}%
\bibitem [{\citenamefont {Dogan}\ \emph {et~al.}(2023)\citenamefont {Dogan},
  \citenamefont {Liou},\ and\ \citenamefont {Chelikowsky}}]{dogan2023}%
  \BibitemOpen
  \bibfield  {author} {\bibinfo {author} {\bibfnamefont {M.}~\bibnamefont
  {Dogan}}, \bibinfo {author} {\bibfnamefont {K.-H.}\ \bibnamefont {Liou}},\
  and\ \bibinfo {author} {\bibfnamefont {J.~R.}\ \bibnamefont {Chelikowsky}},\
  }\bibfield  {title} {\bibinfo {title} {Solving the electronic structure
  problem for over 100 000 atoms in real space},\ }\href
  {https://doi.org/10.1103/PhysRevMaterials.7.L063001} {\bibfield  {journal}
  {\bibinfo  {journal} {Phys. Rev. Mater.}\ }\textbf {\bibinfo {volume} {7}},\
  \bibinfo {pages} {L063001} (\bibinfo {year} {2023})}\BibitemShut {NoStop}%
\bibitem [{\citenamefont {Perdew}\ \emph {et~al.}(1996)\citenamefont {Perdew},
  \citenamefont {Burke},\ and\ \citenamefont {Ernzerhof}}]{perdew1996}%
  \BibitemOpen
  \bibfield  {author} {\bibinfo {author} {\bibfnamefont {J.~P.}\ \bibnamefont
  {Perdew}}, \bibinfo {author} {\bibfnamefont {K.}~\bibnamefont {Burke}},\ and\
  \bibinfo {author} {\bibfnamefont {M.}~\bibnamefont {Ernzerhof}},\ }\bibfield
  {title} {\bibinfo {title} {Generalized gradient approximation made simple},\
  }\href {https://doi.org/10.1103/PhysRevLett.77.3865} {\bibfield  {journal}
  {\bibinfo  {journal} {Phys. Rev. Lett.}\ }\textbf {\bibinfo {volume} {77}},\
  \bibinfo {pages} {3865} (\bibinfo {year} {1996})}\BibitemShut {NoStop}%
\bibitem [{\citenamefont {Troullier}\ and\ \citenamefont
  {Martins}(1991)}]{troullier1991}%
  \BibitemOpen
  \bibfield  {author} {\bibinfo {author} {\bibfnamefont {N.}~\bibnamefont
  {Troullier}}\ and\ \bibinfo {author} {\bibfnamefont {J.~L.}\ \bibnamefont
  {Martins}},\ }\bibfield  {title} {\bibinfo {title} {Efficient
  pseudopotentials for plane-wave calculations},\ }\href
  {https://doi.org/10.1103/PhysRevB.43.1993} {\bibfield  {journal} {\bibinfo
  {journal} {Phys. Rev. B}\ }\textbf {\bibinfo {volume} {43}},\ \bibinfo
  {pages} {1993} (\bibinfo {year} {1991})}\BibitemShut {NoStop}%
\bibitem [{\citenamefont {Weigend}(2002)}]{weigend2002}%
  \BibitemOpen
  \bibfield  {author} {\bibinfo {author} {\bibfnamefont {F.}~\bibnamefont
  {Weigend}},\ }\bibfield  {title} {\bibinfo {title} {A fully direct {RI-HF}
  algorithm: Implementation{,} optimised auxiliary basis sets{,} demonstration
  of accuracy and efficiency},\ }\href {https://doi.org/10.1039/B204199P}
  {\bibfield  {journal} {\bibinfo  {journal} {Phys. Chem. Chem. Phys.}\
  }\textbf {\bibinfo {volume} {4}},\ \bibinfo {pages} {4285} (\bibinfo {year}
  {2002})}\BibitemShut {NoStop}%
\bibitem [{\citenamefont {Ren}\ \emph {et~al.}(2012)\citenamefont {Ren},
  \citenamefont {Rinke}, \citenamefont {Blum}, \citenamefont {Wieferink},
  \citenamefont {Tkatchenko}, \citenamefont {Sanfilippo}, \citenamefont
  {Reuter},\ and\ \citenamefont {Scheffler}}]{Ren_2012}%
  \BibitemOpen
  \bibfield  {author} {\bibinfo {author} {\bibfnamefont {X.}~\bibnamefont
  {Ren}}, \bibinfo {author} {\bibfnamefont {P.}~\bibnamefont {Rinke}}, \bibinfo
  {author} {\bibfnamefont {V.}~\bibnamefont {Blum}}, \bibinfo {author}
  {\bibfnamefont {J.}~\bibnamefont {Wieferink}}, \bibinfo {author}
  {\bibfnamefont {A.}~\bibnamefont {Tkatchenko}}, \bibinfo {author}
  {\bibfnamefont {A.}~\bibnamefont {Sanfilippo}}, \bibinfo {author}
  {\bibfnamefont {K.}~\bibnamefont {Reuter}},\ and\ \bibinfo {author}
  {\bibfnamefont {M.}~\bibnamefont {Scheffler}},\ }\bibfield  {title} {\bibinfo
  {title} {Resolution-of-identity approach to {Hartree-Fock}, hybrid density
  functionals, {RPA}, {MP}2 and {GW} with numeric atom-centered orbital basis
  functions},\ }\href {https://doi.org/10.1088/1367-2630/14/5/053020}
  {\bibfield  {journal} {\bibinfo  {journal} {New Journal of Physics}\ }\textbf
  {\bibinfo {volume} {14}},\ \bibinfo {pages} {053020} (\bibinfo {year}
  {2012})}\BibitemShut {NoStop}%
\bibitem [{\citenamefont {Hu}\ \emph {et~al.}(2020)\citenamefont {Hu},
  \citenamefont {Liu}, \citenamefont {Li}, \citenamefont {Ding}, \citenamefont
  {Yang},\ and\ \citenamefont {Yang}}]{hu2020_ISDF_LRTDDFT}%
  \BibitemOpen
  \bibfield  {author} {\bibinfo {author} {\bibfnamefont {W.}~\bibnamefont
  {Hu}}, \bibinfo {author} {\bibfnamefont {J.}~\bibnamefont {Liu}}, \bibinfo
  {author} {\bibfnamefont {Y.}~\bibnamefont {Li}}, \bibinfo {author}
  {\bibfnamefont {Z.}~\bibnamefont {Ding}}, \bibinfo {author} {\bibfnamefont
  {C.}~\bibnamefont {Yang}},\ and\ \bibinfo {author} {\bibfnamefont
  {J.}~\bibnamefont {Yang}},\ }\bibfield  {title} {\bibinfo {title}
  {Accelerating excitation energy computation in molecules and solids within
  linear-response time-dependent density functional theory via interpolative
  separable density fitting decomposition},\ }\href
  {https://doi.org/10.1021/acs.jctc.9b01019} {\bibfield  {journal} {\bibinfo
  {journal} {Journal of Chemical Theory and Computation}\ }\textbf {\bibinfo
  {volume} {16}},\ \bibinfo {pages} {964} (\bibinfo {year} {2020})}\BibitemShut
  {NoStop}%
\bibitem [{\citenamefont {Lu}\ and\ \citenamefont {Ying}(2015)}]{LU2015329}%
  \BibitemOpen
  \bibfield  {author} {\bibinfo {author} {\bibfnamefont {J.}~\bibnamefont
  {Lu}}\ and\ \bibinfo {author} {\bibfnamefont {L.}~\bibnamefont {Ying}},\
  }\bibfield  {title} {\bibinfo {title} {Compression of the electron repulsion
  integral tensor in tensor hypercontraction format with cubic scaling cost},\
  }\href {https://doi.org/https://doi.org/10.1016/j.jcp.2015.09.014} {\bibfield
   {journal} {\bibinfo  {journal} {Journal of Computational Physics}\ }\textbf
  {\bibinfo {volume} {302}},\ \bibinfo {pages} {329} (\bibinfo {year}
  {2015})}\BibitemShut {NoStop}%
\bibitem [{\citenamefont {Gao}\ and\ \citenamefont
  {Chelikowsky}(2020)}]{gao2020}%
  \BibitemOpen
  \bibfield  {author} {\bibinfo {author} {\bibfnamefont {W.}~\bibnamefont
  {Gao}}\ and\ \bibinfo {author} {\bibfnamefont {J.~R.}\ \bibnamefont
  {Chelikowsky}},\ }\bibfield  {title} {\bibinfo {title} {Accelerating
  time-dependent density functional theory and {GW} calculations for molecules
  and nanoclusters with symmetry adapted interpolative separable density
  fitting},\ }\href {https://doi.org/10.1021/acs.jctc.9b01025} {\bibfield
  {journal} {\bibinfo  {journal} {Journal of Chemical Theory and Computation}\
  }\textbf {\bibinfo {volume} {16}},\ \bibinfo {pages} {2216} (\bibinfo {year}
  {2020})}\BibitemShut {NoStop}%
\bibitem [{\citenamefont {Barbagiovanni}\ \emph {et~al.}(2014)\citenamefont
  {Barbagiovanni}, \citenamefont {Lockwood}, \citenamefont {Simpson},\ and\
  \citenamefont {Goncharova}}]{Barbagiovanni2014}%
  \BibitemOpen
  \bibfield  {author} {\bibinfo {author} {\bibfnamefont {E.~G.}\ \bibnamefont
  {Barbagiovanni}}, \bibinfo {author} {\bibfnamefont {D.~J.}\ \bibnamefont
  {Lockwood}}, \bibinfo {author} {\bibfnamefont {P.~J.}\ \bibnamefont
  {Simpson}},\ and\ \bibinfo {author} {\bibfnamefont {L.~V.}\ \bibnamefont
  {Goncharova}},\ }\bibfield  {title} {\bibinfo {title} {{Quantum confinement
  in Si and Ge nanostructures: Theory and experiment}},\ }\href
  {https://doi.org/10.1063/1.4835095} {\bibfield  {journal} {\bibinfo
  {journal} {Applied Physics Reviews}\ }\textbf {\bibinfo {volume} {1}},\
  \bibinfo {pages} {011302} (\bibinfo {year} {2014})}\BibitemShut {NoStop}%
\bibitem [{\citenamefont {Liang}\ \emph {et~al.}(2019)\citenamefont {Liang},
  \citenamefont {Huang},\ and\ \citenamefont {Gong}}]{liang2019}%
  \BibitemOpen
  \bibfield  {author} {\bibinfo {author} {\bibfnamefont {J.}~\bibnamefont
  {Liang}}, \bibinfo {author} {\bibfnamefont {C.}~\bibnamefont {Huang}},\ and\
  \bibinfo {author} {\bibfnamefont {X.}~\bibnamefont {Gong}},\ }\bibfield
  {title} {\bibinfo {title} {Silicon nanocrystals and their composites:
  Syntheses, fluorescence mechanisms, and biological applications},\ }\href
  {https://doi.org/10.1021/acssuschemeng.9b04359} {\bibfield  {journal}
  {\bibinfo  {journal} {ACS Sustainable Chemistry \& Engineering}\ }\textbf
  {\bibinfo {volume} {7}},\ \bibinfo {pages} {18213} (\bibinfo {year}
  {2019})}\BibitemShut {NoStop}%
\bibitem [{\citenamefont {Dohnalov{\'a}}\ \emph {et~al.}(2013)\citenamefont
  {Dohnalov{\'a}}, \citenamefont {Poddubny}, \citenamefont {Prokofiev},
  \citenamefont {de~Boer}, \citenamefont {Umesh}, \citenamefont {Paulusse},
  \citenamefont {Zuilhof},\ and\ \citenamefont
  {Gregorkiewicz}}]{Dohnalová2013}%
  \BibitemOpen
  \bibfield  {author} {\bibinfo {author} {\bibfnamefont {K.}~\bibnamefont
  {Dohnalov{\'a}}}, \bibinfo {author} {\bibfnamefont {A.~N.}\ \bibnamefont
  {Poddubny}}, \bibinfo {author} {\bibfnamefont {A.~A.}\ \bibnamefont
  {Prokofiev}}, \bibinfo {author} {\bibfnamefont {W.~D.}\ \bibnamefont
  {de~Boer}}, \bibinfo {author} {\bibfnamefont {C.~P.}\ \bibnamefont {Umesh}},
  \bibinfo {author} {\bibfnamefont {J.~M.}\ \bibnamefont {Paulusse}}, \bibinfo
  {author} {\bibfnamefont {H.}~\bibnamefont {Zuilhof}},\ and\ \bibinfo {author}
  {\bibfnamefont {T.}~\bibnamefont {Gregorkiewicz}},\ }\bibfield  {title}
  {\bibinfo {title} {Surface brightens up {Si} quantum dots: direct
  bandgap-like size-tunable emission},\ }\href
  {https://doi.org/10.1038/lsa.2013.3} {\bibfield  {journal} {\bibinfo
  {journal} {Light: Science {\&} Applications}\ }\textbf {\bibinfo {volume}
  {2}},\ \bibinfo {pages} {e47} (\bibinfo {year} {2013})}\BibitemShut {NoStop}%
\bibitem [{\citenamefont {Neiner}\ \emph {et~al.}(2006)\citenamefont {Neiner},
  \citenamefont {Chiu},\ and\ \citenamefont {Kauzlarich}}]{neiner2006}%
  \BibitemOpen
  \bibfield  {author} {\bibinfo {author} {\bibfnamefont {D.}~\bibnamefont
  {Neiner}}, \bibinfo {author} {\bibfnamefont {H.~W.}\ \bibnamefont {Chiu}},\
  and\ \bibinfo {author} {\bibfnamefont {S.~M.}\ \bibnamefont {Kauzlarich}},\
  }\bibfield  {title} {\bibinfo {title} {Low-temperature solution route to
  macroscopic amounts of hydrogen terminated silicon nanoparticles},\ }\href
  {https://doi.org/10.1021/ja064177q} {\bibfield  {journal} {\bibinfo
  {journal} {Journal of the American Chemical Society}\ }\textbf {\bibinfo
  {volume} {128}},\ \bibinfo {pages} {11016} (\bibinfo {year}
  {2006})}\BibitemShut {NoStop}%
\bibitem [{\citenamefont {Huan}\ and\ \citenamefont
  {Shu-Qing}(2014)}]{Chang_2014}%
  \BibitemOpen
  \bibfield  {author} {\bibinfo {author} {\bibfnamefont {C.}~\bibnamefont
  {Huan}}\ and\ \bibinfo {author} {\bibfnamefont {S.}~\bibnamefont
  {Shu-Qing}},\ }\bibfield  {title} {\bibinfo {title} {Silicon nanoparticles:
  Preparation, properties, and applications},\ }\href
  {https://doi.org/10.1088/1674-1056/23/8/088102} {\bibfield  {journal}
  {\bibinfo  {journal} {Chinese Physics B}\ }\textbf {\bibinfo {volume} {23}},\
  \bibinfo {pages} {088102} (\bibinfo {year} {2014})}\BibitemShut {NoStop}%
\bibitem [{\citenamefont {Godefroo}\ \emph {et~al.}(2008)\citenamefont
  {Godefroo}, \citenamefont {Hayne}, \citenamefont {Jivanescu}, \citenamefont
  {Stesmans}, \citenamefont {Zacharias}, \citenamefont {Lebedev}, \citenamefont
  {Van~Tendeloo},\ and\ \citenamefont {Moshchalkov}}]{Godefroo2008}%
  \BibitemOpen
  \bibfield  {author} {\bibinfo {author} {\bibfnamefont {S.}~\bibnamefont
  {Godefroo}}, \bibinfo {author} {\bibfnamefont {M.}~\bibnamefont {Hayne}},
  \bibinfo {author} {\bibfnamefont {M.}~\bibnamefont {Jivanescu}}, \bibinfo
  {author} {\bibfnamefont {A.}~\bibnamefont {Stesmans}}, \bibinfo {author}
  {\bibfnamefont {M.}~\bibnamefont {Zacharias}}, \bibinfo {author}
  {\bibfnamefont {O.~I.}\ \bibnamefont {Lebedev}}, \bibinfo {author}
  {\bibfnamefont {G.}~\bibnamefont {Van~Tendeloo}},\ and\ \bibinfo {author}
  {\bibfnamefont {V.~V.}\ \bibnamefont {Moshchalkov}},\ }\bibfield  {title}
  {\bibinfo {title} {Classification and control of the origin of
  photoluminescence from {Si} nanocrystals},\ }\href
  {https://doi.org/10.1038/nnano.2008.7} {\bibfield  {journal} {\bibinfo
  {journal} {Nature Nanotechnology}\ }\textbf {\bibinfo {volume} {3}},\
  \bibinfo {pages} {174} (\bibinfo {year} {2008})}\BibitemShut {NoStop}%
\bibitem [{\citenamefont {Priolo}\ \emph {et~al.}(2014)\citenamefont {Priolo},
  \citenamefont {Gregorkiewicz}, \citenamefont {Galli},\ and\ \citenamefont
  {Krauss}}]{Priolo2014}%
  \BibitemOpen
  \bibfield  {author} {\bibinfo {author} {\bibfnamefont {F.}~\bibnamefont
  {Priolo}}, \bibinfo {author} {\bibfnamefont {T.}~\bibnamefont
  {Gregorkiewicz}}, \bibinfo {author} {\bibfnamefont {M.}~\bibnamefont
  {Galli}},\ and\ \bibinfo {author} {\bibfnamefont {T.~F.}\ \bibnamefont
  {Krauss}},\ }\bibfield  {title} {\bibinfo {title} {Silicon nanostructures for
  photonics and photovoltaics},\ }\href
  {https://doi.org/10.1038/nnano.2013.271} {\bibfield  {journal} {\bibinfo
  {journal} {Nature Nanotechnology}\ }\textbf {\bibinfo {volume} {9}},\
  \bibinfo {pages} {19} (\bibinfo {year} {2014})}\BibitemShut {NoStop}%
\bibitem [{\citenamefont {Liao}\ \emph {et~al.}(2022)\citenamefont {Liao},
  \citenamefont {Liou},\ and\ \citenamefont {Chelikowsky}}]{Liao2022}%
  \BibitemOpen
  \bibfield  {author} {\bibinfo {author} {\bibfnamefont {T.}~\bibnamefont
  {Liao}}, \bibinfo {author} {\bibfnamefont {K.-H.}\ \bibnamefont {Liou}},\
  and\ \bibinfo {author} {\bibfnamefont {J.~R.}\ \bibnamefont {Chelikowsky}},\
  }\bibfield  {title} {\bibinfo {title} {Dielectric screening and vacancy
  formation for large neutral and charged {Si}$_{n}${H}$_{m}$ $(n > 1500)$
  nanocrystals using real-space pseudopotentials},\ }\href
  {https://doi.org/10.1103/PhysRevMaterials.6.054603} {\bibfield  {journal}
  {\bibinfo  {journal} {Phys. Rev. Mater.}\ }\textbf {\bibinfo {volume} {6}},\
  \bibinfo {pages} {054603} (\bibinfo {year} {2022})}\BibitemShut {NoStop}%
\bibitem [{\citenamefont {Neuhauser}\ \emph {et~al.}(2014)\citenamefont
  {Neuhauser}, \citenamefont {Gao}, \citenamefont {Arntsen}, \citenamefont
  {Karshenas}, \citenamefont {Rabani},\ and\ \citenamefont
  {Baer}}]{Neuhauser2014}%
  \BibitemOpen
  \bibfield  {author} {\bibinfo {author} {\bibfnamefont {D.}~\bibnamefont
  {Neuhauser}}, \bibinfo {author} {\bibfnamefont {Y.}~\bibnamefont {Gao}},
  \bibinfo {author} {\bibfnamefont {C.}~\bibnamefont {Arntsen}}, \bibinfo
  {author} {\bibfnamefont {C.}~\bibnamefont {Karshenas}}, \bibinfo {author}
  {\bibfnamefont {E.}~\bibnamefont {Rabani}},\ and\ \bibinfo {author}
  {\bibfnamefont {R.}~\bibnamefont {Baer}},\ }\bibfield  {title} {\bibinfo
  {title} {Breaking the theoretical scaling limit for predicting quasiparticle
  energies: The stochastic {GW} approach},\ }\href
  {https://doi.org/10.1103/PhysRevLett.113.076402} {\bibfield  {journal}
  {\bibinfo  {journal} {Phys. Rev. Lett.}\ }\textbf {\bibinfo {volume} {113}},\
  \bibinfo {pages} {076402} (\bibinfo {year} {2014})}\BibitemShut {NoStop}%
\bibitem [{\citenamefont {Brabec}\ \emph {et~al.}(2015)\citenamefont {Brabec},
  \citenamefont {Lin}, \citenamefont {Shao}, \citenamefont {Govind},
  \citenamefont {Yang}, \citenamefont {Saad},\ and\ \citenamefont
  {Ng}}]{brabec2015}%
  \BibitemOpen
  \bibfield  {author} {\bibinfo {author} {\bibfnamefont {J.}~\bibnamefont
  {Brabec}}, \bibinfo {author} {\bibfnamefont {L.}~\bibnamefont {Lin}},
  \bibinfo {author} {\bibfnamefont {M.}~\bibnamefont {Shao}}, \bibinfo {author}
  {\bibfnamefont {N.}~\bibnamefont {Govind}}, \bibinfo {author} {\bibfnamefont
  {C.}~\bibnamefont {Yang}}, \bibinfo {author} {\bibfnamefont {Y.}~\bibnamefont
  {Saad}},\ and\ \bibinfo {author} {\bibfnamefont {E.~G.}\ \bibnamefont {Ng}},\
  }\bibfield  {title} {\bibinfo {title} {Efficient algorithms for estimating
  the absorption spectrum within linear response {TDDFT}},\ }\href
  {https://doi.org/10.1021/acs.jctc.5b00887} {\bibfield  {journal} {\bibinfo
  {journal} {Journal of Chemical Theory and Computation}\ }\textbf {\bibinfo
  {volume} {11}},\ \bibinfo {pages} {5197} (\bibinfo {year}
  {2015})}\BibitemShut {NoStop}%
\end{thebibliography}%



\begin{figure*} 
 \includegraphics[width=1.0\linewidth]{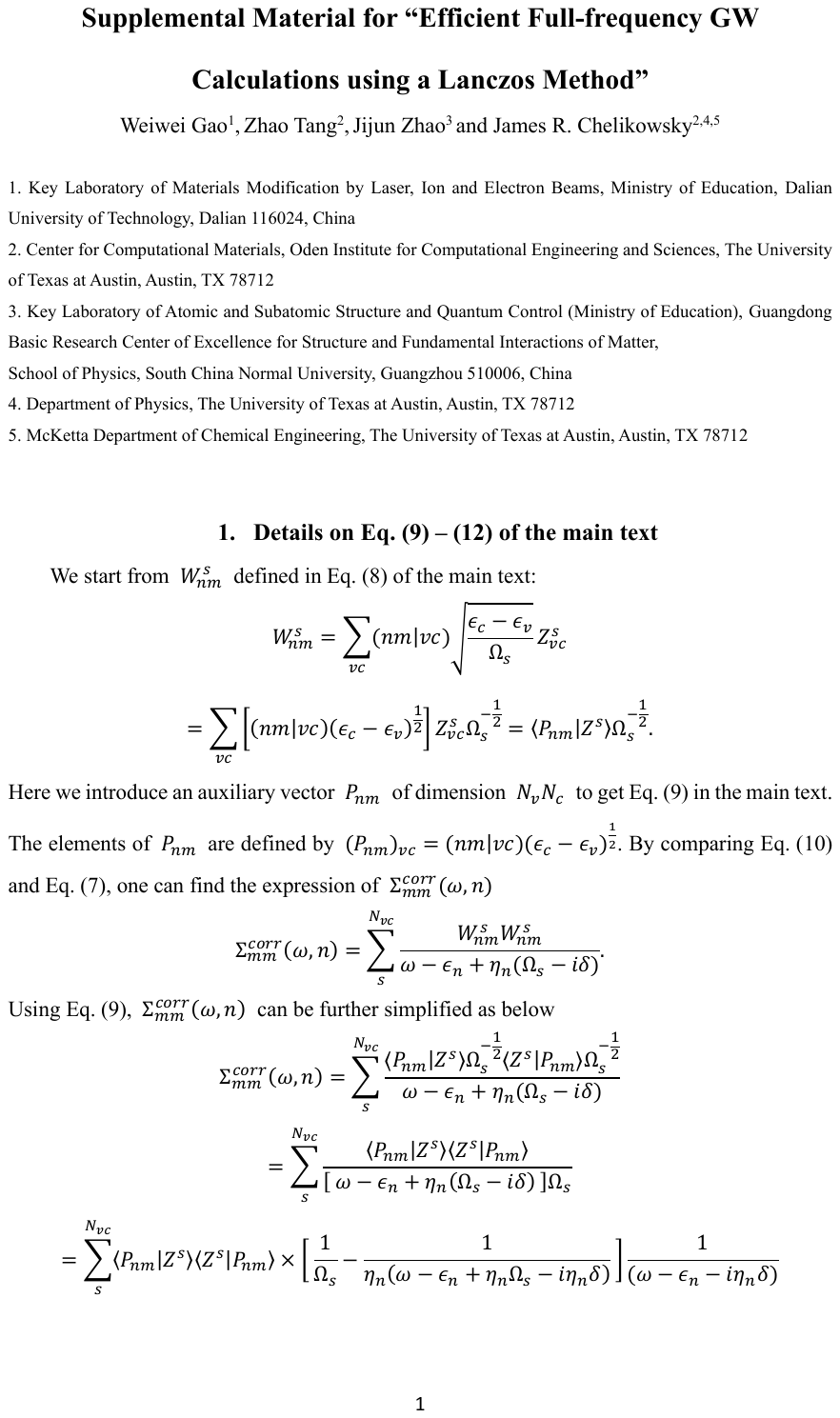}
\end{figure*}
\begin{figure*} 
 \includegraphics[width=1.0\linewidth]{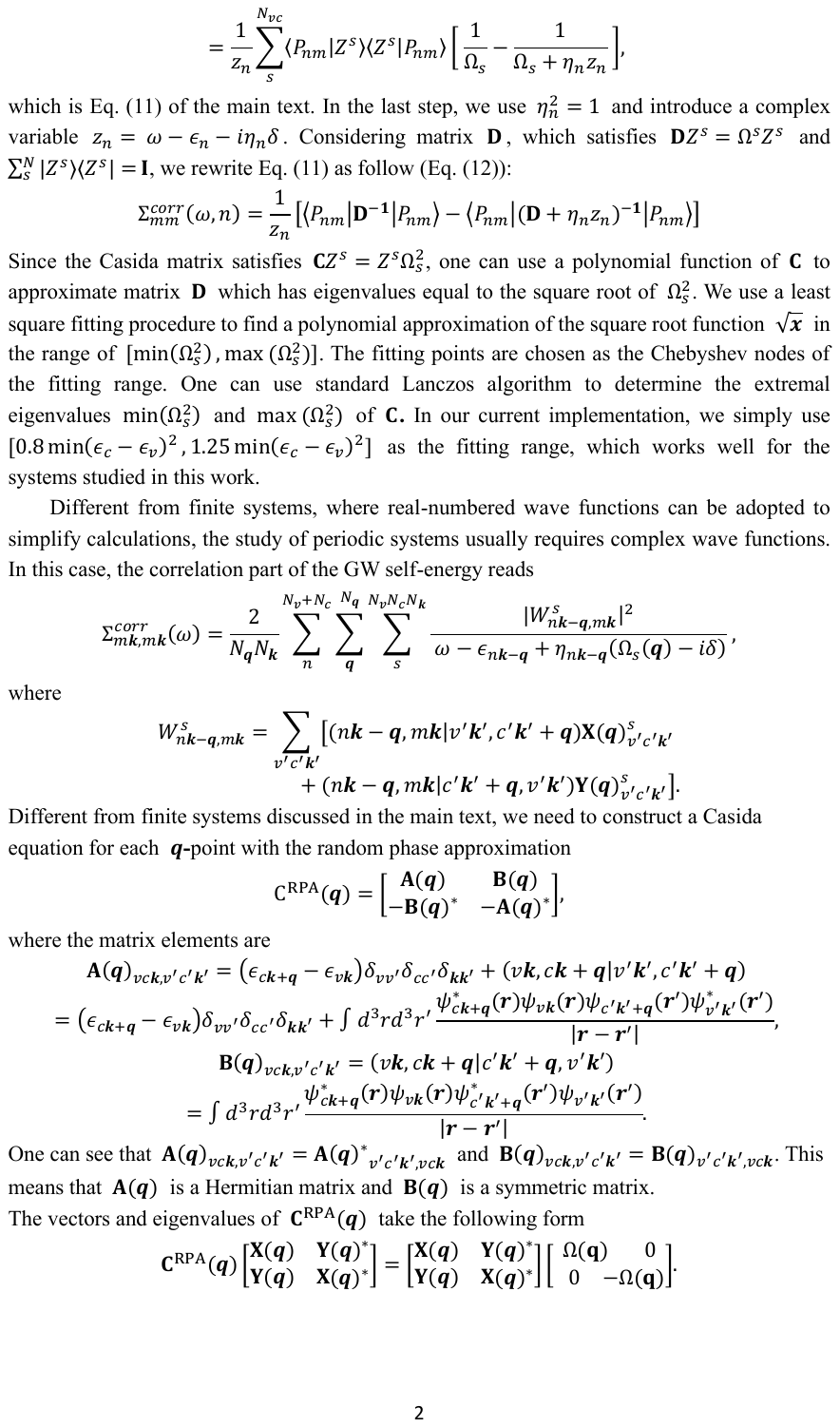}
\end{figure*}
\begin{figure*} 
 \includegraphics[width=1.0\linewidth]{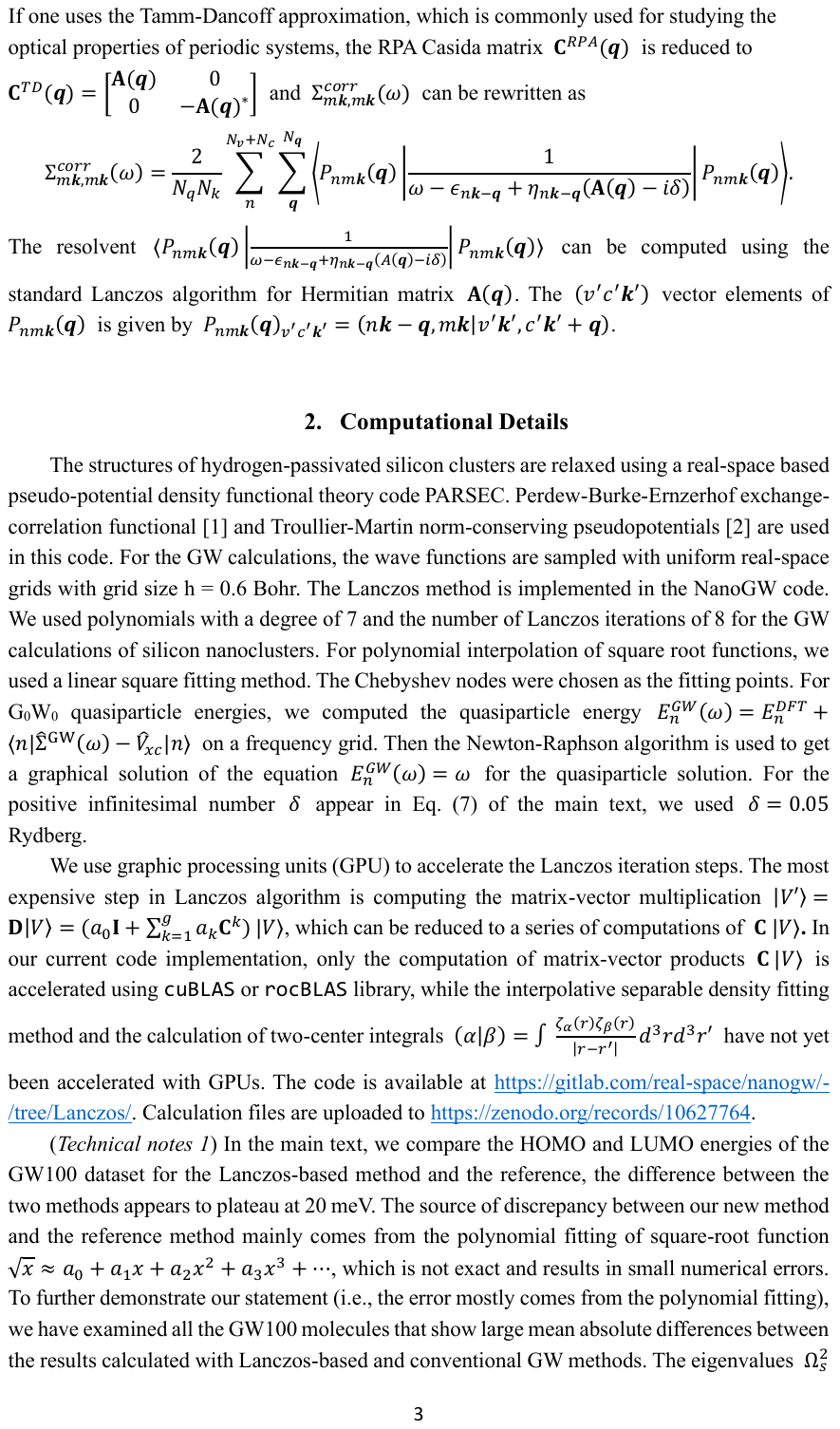}
\end{figure*}
\begin{figure*} 
 \includegraphics[width=1.0\linewidth]{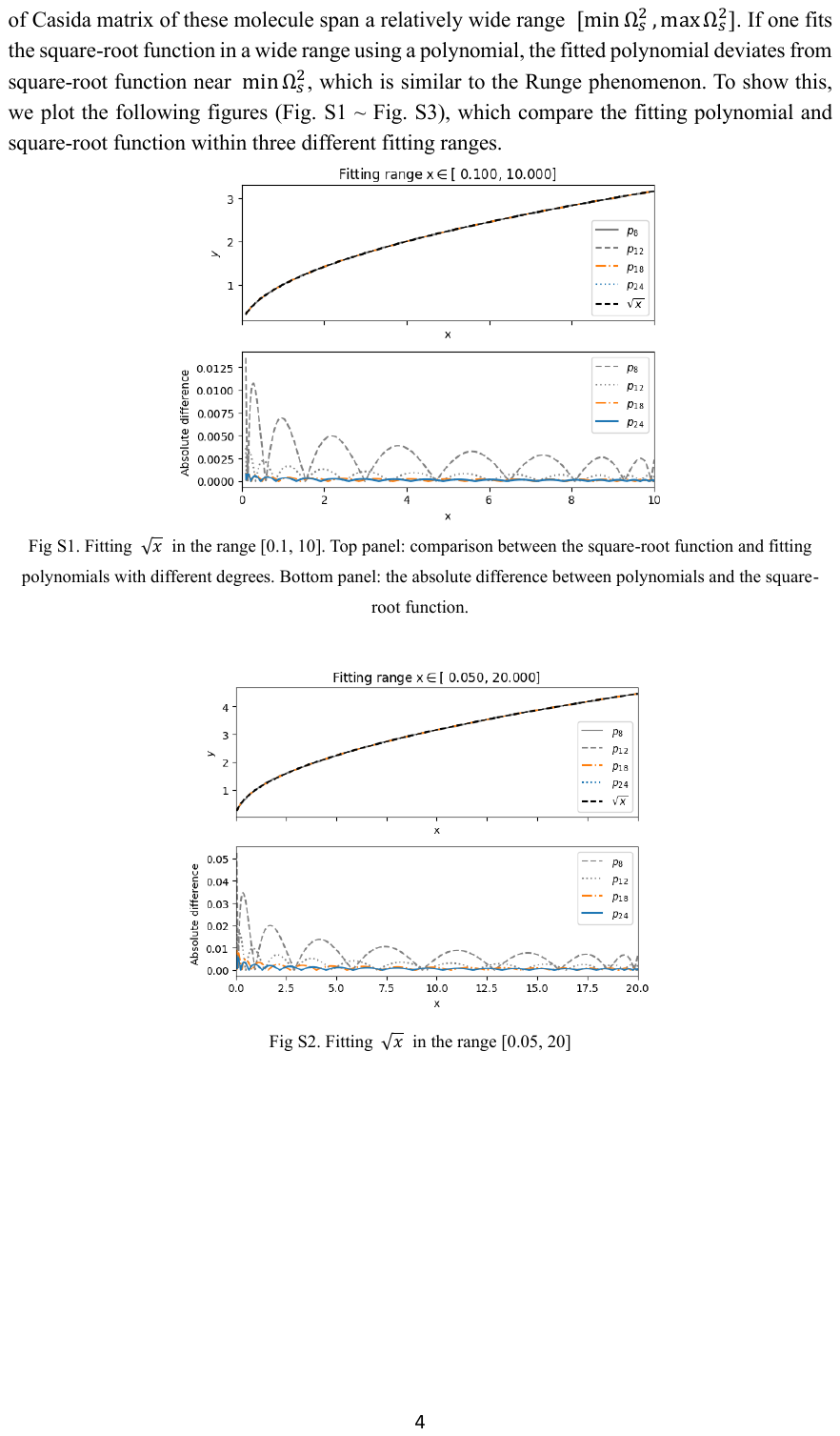}
\end{figure*}
\begin{figure*} 
 \includegraphics[width=1.0\linewidth]{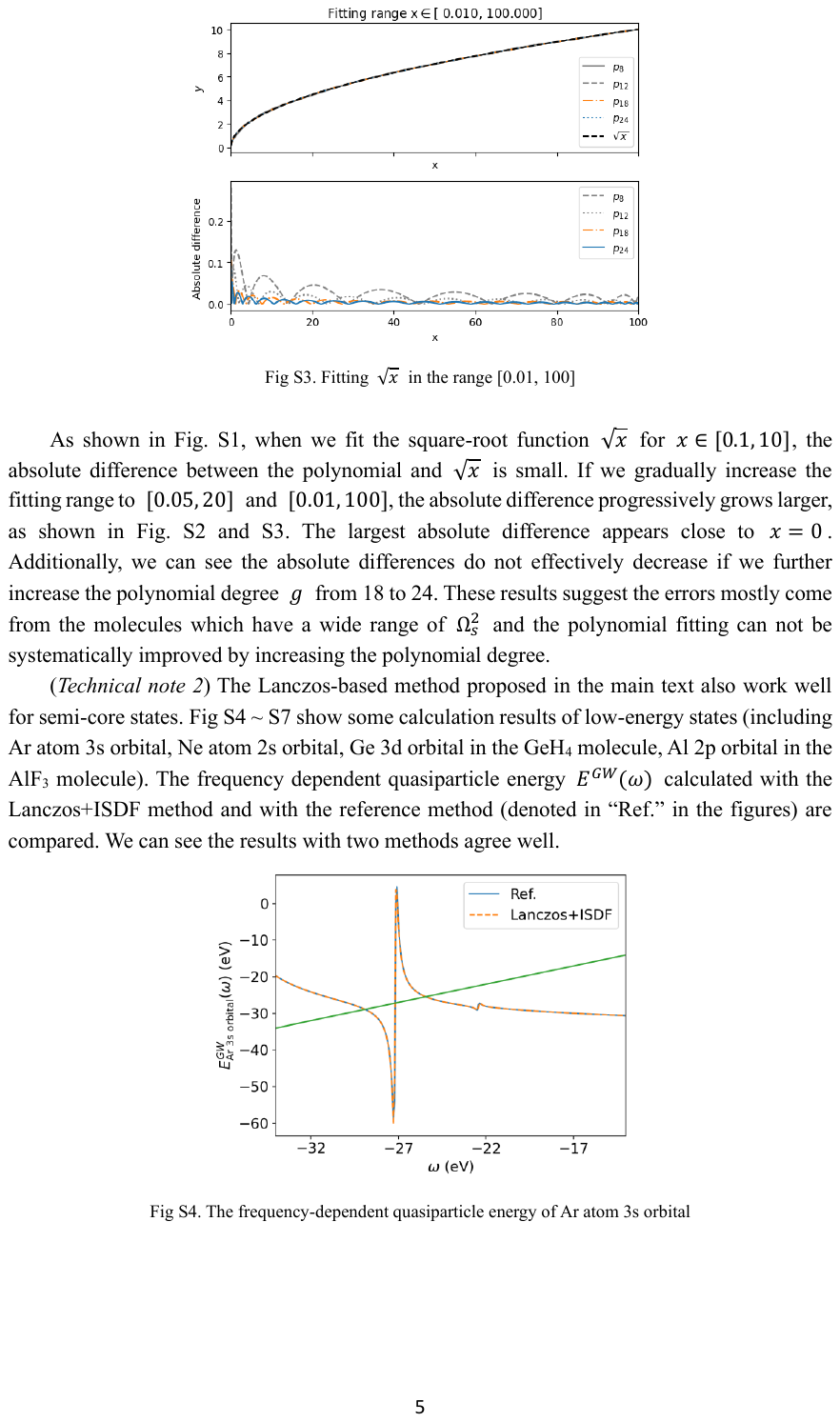}
\end{figure*}
\begin{figure*} 
 \includegraphics[width=1.0\linewidth]{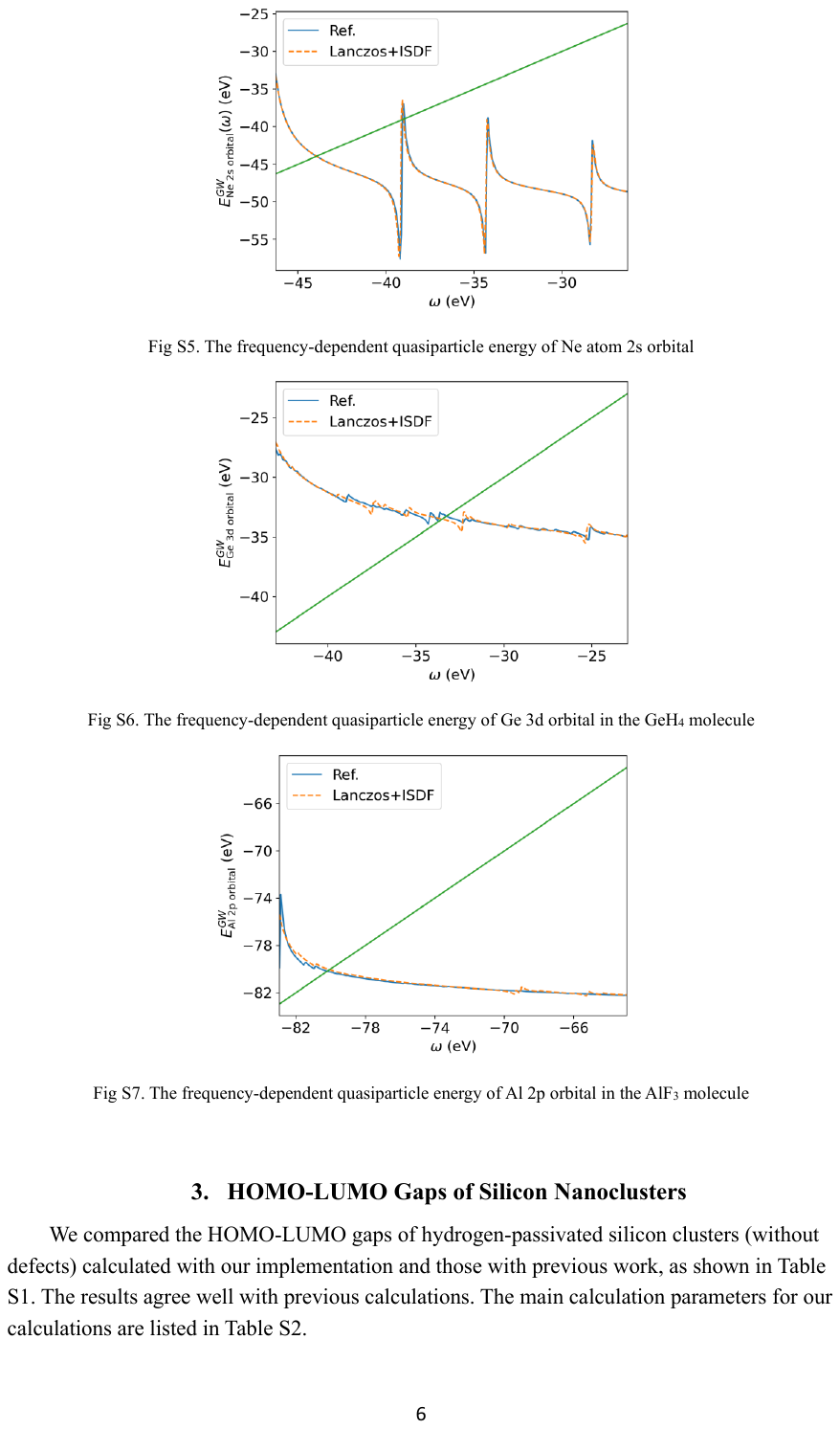}
\end{figure*}
\begin{figure*} 
 \includegraphics[width=1.0\linewidth]{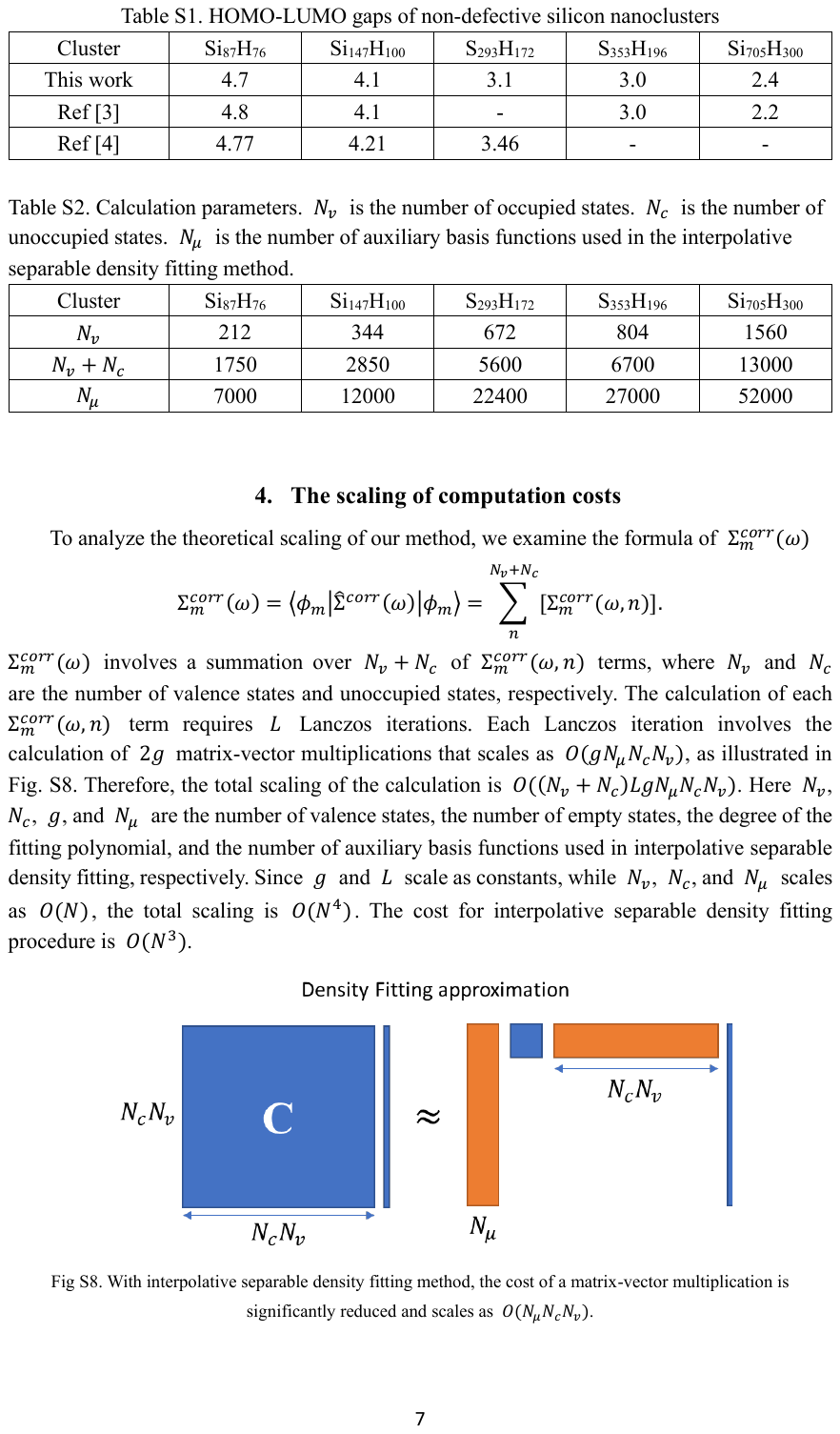}
\end{figure*}
\begin{figure*} 
 \includegraphics[width=1.0\linewidth]{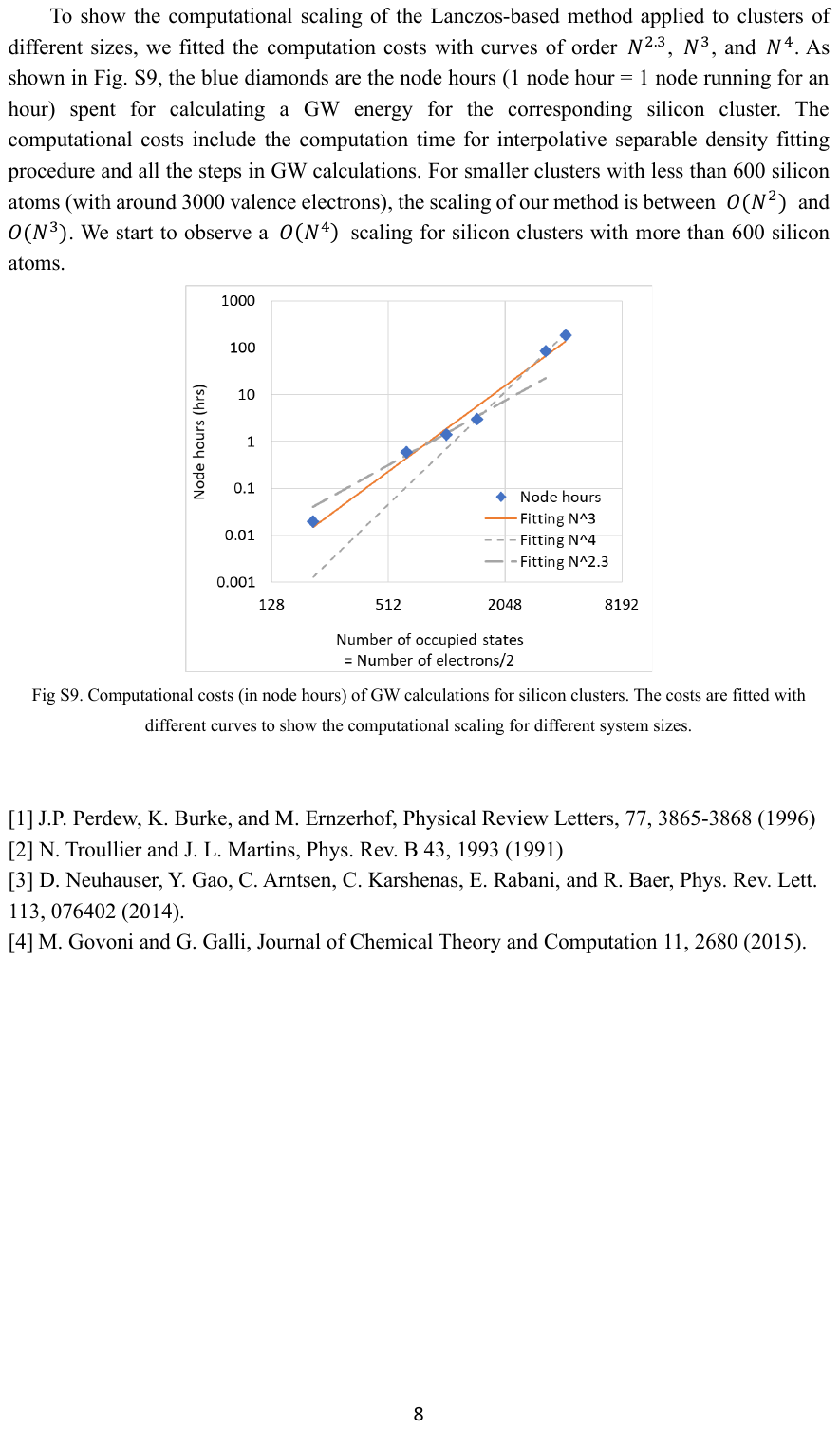}
\end{figure*}

\end{document}